\documentclass[a4paper,11pt]{article}
\usepackage{amssymb,amsmath,amsthm}
\usepackage{bm}%
\usepackage{ascmac}
\usepackage{mathrsfs}
\usepackage{stmaryrd}
\usepackage{comment}
\usepackage{multirow}
\usepackage{braket}
\usepackage{tikz}
    \usetikzlibrary{arrows.meta,calc}
\usepackage{xcolor}
\usepackage{hyperref}
\hypersetup{
	colorlinks=true,
	citecolor=blue,
	linkcolor=blue,
	urlcolor=blue,
}

\def\nn{\notag}
\def\Z2{\mathbb{Z}_2^2}
\def\g{\mathfrak{g}}
\def\h{\mathfrak{h}}

\def\DP#1#2{\hat{#1}\cdot \hat{#2}}
\def\PH#1#2{(-1)^{\hat{#1}\cdot \hat{#2}}}

\def\osp{\mathfrak{osp}}
\def\sl{\mathfrak{sl}}

\numberwithin{equation}{section} 

\title{A Twisted Affine Interpolation of the $\mathbb{Z}_2^2$-graded Integrable Super-Liouville and super-Sinh-Gordon Equations}
\author{Naruhiko Aizawa${}^{1}$\thanks{{E-mail: {\it aizawa@omu.ac.jp}}}, \quad 
	 \quad
	Ren Ito${}^{1}$\thanks{E-mail: \textit{sd22709y@st.omu.ac.jp}}, \\[5pt]
	Luiza Miranda${}^{2}$\thanks{E-mail: \textit{luiza16@cbpf.br}} \quad and\quad
	Francesco Toppan${}^{2}$\thanks{{E-mail: {\it toppan@cbpf.br}}}
}
\date{\today}

\begin{document}
	
	\maketitle
	\thispagestyle{empty}
	\begin{center}
		{\small{	${}^{1}$\textit{Department of Physics, Graduate School of Science,
					\\
					Osaka Metropolitan University, Sugimoto Campus,
					\\
					Osaka 558-8585, Japan.}
					
				\bigskip
				${}^{2}$\textit{
					CBPF, Rua Dr. Xavier Sigaud 150, Urca,
					\\
					cep 22290-180, Rio de Janeiro (RJ), Brazil.
				}
		}}
	\end{center}

\vfill	
\begin{abstract}
This paper is a third one in a series devoted to the construction of $2D$, $\mathbb{Z}_2^2$-graded, integrable Toda field theories. The first paper introduced parabosonic models obtained from Rittenberg-Wyler color Lie algebras; the second paper explained how to deal with parafermions (the $\mathbb{Z}_2^2$-graded super-Liouville theory) from finite Rittenberg-Wyler color Lie superalgebras. This new paper extends the construction to the infinite case of (twisted) affine color Lie superalgebras. A $\mathbb{Z}_2^2$-graded extension of ${\widehat{\mathfrak{osp}(2|2)^{(2)}}}$ allows to interpolate the super-Liouville and the super-Sinh-Gordon models, recovered as special limits. The $\mathbb{Z}_2^2$-graded extensions of three frameworks are here discussed: the Polyakov's soldering, the Leznov-Savel'ev Lax pair (zero-curvature) formulation and the B\"acklund transformations. As a byproduct of the construction, the Hamiltonian reduction from the $\mathbb{Z}_2^2$-graded ${\widehat{\mathfrak{osp}(2|2)^{(2)}}}$ WZNW models induces a $\mathbb{Z}_2^2$-graded color Lie superalgebra which admits three inequivalent superVirasoro subalgebras; they are the ${\cal N}=2$ superVirasoro, the $(1,\frac{3}{2},\frac{3}{2},2)$ $\mathbb{Z}_2^2$-graded superVirasoro and the 	$(\frac{3}{2},\frac{3}{2},2,2)$ $\mathbb{Z}_2^2$-graded superVirasoro.
\end{abstract}

\clearpage
\setcounter{page}{1}

\section{Introduction}

This paper is the third one in a series devoted to the systematic construction of two-dimensional, $\mathbb{Z}_2^2$-graded, integrable Toda field theories.  The first paper \cite{aiktt} introduced parabosonic models obtained from Rittenberg-Wyler \cite{{rw1},{rw2}} color Lie algebras; it presented integrable $\mathbb{Z}_2^2$-graded versions of the Liouville and of the Sinh-Gordon models from, respectively,  $\mathbb{Z}_2^2$-graded color extensions of the $\sl(2)$ and affine 
${\widehat{\sl(2)}}$ algebras. The second paper \cite{afitt} discussed the challenge of introducing parafermions from Rittenberg-Wyler color Lie superalgebras with a finite number of generators; in that work the $\mathbb{Z}_2^2$-graded super-Liouville theory was obtained from a $\mathbb{Z}_2^2$-graded extension of $\osp(1|2)$. 
This new paper discusses the further challenge of constructing integrable $\mathbb{Z}_2^2$-graded super-Toda models from infinitely generated $\mathbb{Z}_2^2$-graded color Lie superalgebras.

Let us recall that conformal affine Toda theory associated with $\widehat{\sl(2)}$ reduces to either Liouville theory or sinh-Gordon theory under certain limiting conditions \cite{babo}. 
A supersymmetric counterpart of this bosonic theory is obtained from the twisted affine superalgebra $\widehat{\osp(2|2)^{(2)}}$ \cite{ToppanZhang} (see also \cite{ChaiKu,LeZSavLei,Arvis,ivkr,ToppanSL} for supersymmetric Liouville and sinh-Gordon equations). 
It is then natural to consider a $\Z2$-graded extension of this theory based on a $\Z2$-graded $\widehat{\osp(2|2)^{(2)}}$.  
We construct the Lax operator of this new theory using the finite dimensional $\Z2$-graded  color $\osp(2|2)$ subalgebra. 
We find that  our model recovers, by suitably restricting the parameters, both the $\mathbb{Z}_2^2$-graded superLiouville theory and a  $\mathbb{Z}_2^2$-graded 
super-Sinh-Gordon models. This demonstrates  that the  $\mathbb{Z}_2^2$-graded extension of ${\widehat{\osp(2|2)^{(2)}}}$ allows to interpolate these two models.

The present model, by construction, consists of paraparticles associated with the $\Z2$-graded color Lie superalgebra \cite{yaji,tol1,StoVDJ,StoVDJ3,top2,top1,bfrt}. 
We emphasizes the non-trivial significance of taking such paraparticles into account. 
It has been shown, see \cite{{top1},{top2}} and also \cite{{bfrt},{top3}}, that certain results implied by $\Z2$-graded paraparticles  cannot be reproduced by ordinary Bose-Fermi statistics. 
$\Z2$-graded paraparticles belong to a class of non-trivial parastatistics, inequivalent to either fermions or bosons, which are shown to appear in theoretical condensed matter systems \cite{WaHa}.  
On the experimental side, the possibility of manipulating parastatistics either in simulations \cite{parasim} or in laborator by using trapped ions \cite{paraexp} have been achieved.

The analysis of the integrability of our model is conducted in three different frameworks:
\\
{\it $~~$i}) an extension of the Polyakov's soldering procedure \cite{polyak},\\
{\it $~$ii}) an extension of the Leznov Savel'ev \cite{LezSav1,LezSav2,LezSav3} Lax pair (zero-curvature) formalism and\\
{\it iii}) the B\"acklund transformations.

As a byproduct of the construction we obtain a Hamiltonian reduction from a  $\mathbb{Z}_2^2$-graded $\widehat{\osp(2|2)^{(2)}}$ WZNW model. The Hamiltonian reduction induces an infinitely generated  $\mathbb{Z}_2^2$-graded color Lie superalgebra defined by $8$ fields. It presents three inequivalent super-Virasoro subalgebras given by:\\
$1$)  the ``ordinary" ${\cal N}=2$ super-Virasoro algebra,\\
$2$)  the  $\mathbb{Z}_2^2$-graded super-Virasoro algebra recovered in \cite{afitt} with component fields of scaling dimension $(\frac{3}{2},\frac{3}{2},2,2)$ and \\
$3$) the $\mathbb{Z}_2^2$-graded super-Virasoro algebra introduced in \cite{zhe}  with 
component fields of scaling dimensions  $(1,\frac{3}{2},\frac{3}{2},2)$.

The interpretation of this result goes as follows. As recalled in  \cite{ToppanZhang} a fermionic version of the \cite{LezSav1} Lax pair requires a superalgebra with a Dynkin's diagram presentation in terms of fermionic simple roots. The twisted affine superalgebra $\widehat{\osp(2|2)^{(2)}}$ is the minimal (rank $2$) infinite-dimensional superalgebra satisfying this property, see Table $10$ in \cite{fss}. Its finite dimensional subalgebra $\osp(2|2)$ is isomorphic, see \cite{vritt}, to $\sl(2|1)$; it admits a complex structure and therefore implies \cite{ivto} an ${\cal N}=2$ supersymmetry. This explains the appearance, in the Hamiltonian reduction, of the ${\cal N}=2$ superVirasoro subalgebra.  The $\mathbb{Z}_2^2$-graded superalgebra 
$\osp(1|2)$, being a subalgebra of $\mathbb{Z}_2^2$-graded  ${\widehat{\osp(2|2)^{(2)}}}$, explains the presence of the \cite{afitt} $\mathbb{Z}_2^2$-graded superVirasoro subalgebra. The presence of the extra component fields accommodated in
$\mathbb{Z}_2^2$-graded  ${\widehat{\osp(2|2)^{(2)}}}$ allow to recover the \cite{zhe} $\mathbb{Z}_2^2$-graded super-Virasoro subalgebra.

As another byproduct, we obtain a Lax operator formulation based on the $\Z2$-graded  $\osp(2|2)$  for the lowest-order equations of the $\Z2$-graded sine-Gordon model introduced in \cite{brusigma,bruSG}, which is the first integrable system to be extended to the $\Z2$-graded setting.

Before proceeding with our analysis, we briefly review previous works on  $\Z2$-graded extensions of integrable system. A common approach to constructing integrable systems is to develop integrable hierarchies based on underlying algebraic structures.  In \cite{AiFujiIto}, a $\Z2$-graded $\osp(1|2)$ is employed to construct a hierarchy which includes a $\mathcal{N} = 1 $ $\Z2$-graded extension of the Liouville, sinh-Gordon, cosh-Gordon and modified KdV equations.  
The $\Z2$-graded KdV equation is also derive from the $\Z2$-graded modified KdV equation via a Miura transformation. 
The same algebraic structure is also employed to derive a $\Z2$-graded Camassa-Holm equation \cite{AiFujiItoVas} where its bi-Hamiltonian structure is established. 
In \cite{yuan2026super}, the $\Z2$-graded color Lie superalgebra $\mathfrak{sp}_1(6)$, which is a $\Z2$-graded superextension of $\mathfrak{sp}(6)$ \cite{stvdjclass}, is employed to construct $\Z2$-graded integrable hierarchies in both $(1+1)$ and $(2+1)$ dimensions. 
In contrast, in $(3+1)$-dimensional spacetime, a minimal $\Z2$-graded supersymmetric Yang–Mills Lagrangian was constructed recently \cite{ItoNagoTani} (see also \cite{ItoNago}). It was shown that $\Z2$-graded generalized statistics can be consistently realized at the classical level in a stable, interacting supersymmetric gauge theory. 
This suggests the possibility of deriving the $\Z2$-graded soliton equations obtained so far as symmetry reductions of a suitably formulated anti-selfdual Yang–Mills theory with $\Z2$-graded supersymmetry.

This paper is organized as follows. In the next section, we define a $\Z2$-graded color Lie superalgebra and fix our notation and conventions. 
First, in \S \ref{SEC:Equatoion},  we derive the $\mathcal{N} = 2$ extended $\Z2$-graded super-Liouville equation  via Polyakov's soldering procedure.  
It will be shown in \S \ref{SEC:CurrentAlg} that the Hamiltonian reduction for the associated current algebra, supplemented by an additional gauge fxing inspired by \cite{sab1,sab2}, gives
rise to a $\Z2$-graded extension of the $\mathcal{N} = 2$ super-Virasoro algebra. 
We present its algebraic relations explicitly, thereby demonstrating that this algebra contains three inequivalent super-Virasoro subalgebras. 
Then, in \S \ref{sec:zc-eq},  Lax operator (zero-curvature) formulation based on the $\Z2$-graded $\osp(2|2)$ is developed. We introduce four $\Z2$-graded superfield and show that the zero-curvature formulation gives rise to a system of equations and two constraints. 
We then demonstrate that the $\Z2$-graded super-Liouville equation obtained by the soldering procedure is recovered by an appropriate reduction, while different reduction gives $\Z2$-graded super-sinh-Gordon equation. 
The corresponding B\"acklund transformation for both equations are also presented in \S \ref{sec:zc-eq}.  
Finally, in \S \ref{sec:Z22-sineGordon}, we present a Lax operator formulation of the lowest-order members of the $\Z2$-sine-Gordon equation introduced by Bruce. This is obtained by employing the different superspace from the one used fot $\Z2$-graded super-Liouville and super-sinh-Gordon equations. 
A summary of the present work, together with further remarks and possible directions for
future research, is given in the Conclusion.

\section{$\Z2$-graded extension of $\osp(2|2)$} \label{SEC:alg}
\setcounter{equation}{0}

Let us recall the definition of the $\Z2$-graded color Lie superalgebra \cite{rw1,rw2} (see also \cite{Ree,sch}). 
Let $\mathfrak{g}$ be a vector space and $\hat{a}\equiv [a_1a_2]$ an element of $\Z2$. 
Suppose that $\mathfrak{g}$ is a direct sum of graded components 
\begin{equation}
	\mathfrak{g} = \bigoplus_{\hat{a} \in \Z2}  g_{\hat{a}}= g_{[00]} \oplus g_{[10]} \oplus g_{[01]} \oplus g_{[11]}.   
\end{equation}
If $\mathfrak{g}$ admits a bilinear operation (the graded Lie bracket), denoted by $ \llbracket \cdot, \cdot \rrbracket $ and satisfying the identities 
\begin{align}
	& \llbracket A_{\hat{a}}, B_{\hat{b}}  \rrbracket \in g_{\hat{a}+\hat{b}},
	\\
	& \llbracket A_{\hat{a}}, B_{\hat{b}} \rrbracket = -(-1)^{\hat{a}\cdot\hat{b}} \llbracket  B_{\hat{b}}, A_{\hat{a}} \rrbracket,
	\\
	& (-1)^{\hat{a}\cdot\hat{c}} \llbracket A_{\hat{a}}, \llbracket B_{\hat{b}}, C_{\hat{c}} \rrbracket \rrbracket + \PH{b}{a} \llbracket B_{\hat{b}}, \llbracket C_{\hat{c}}, A_{\hat{a}} \rrbracket \rrbracket + \PH{c}{b}\llbracket C_{\hat{c}}, \llbracket A_{\hat{a}}, B_{\hat{b}} \rrbracket \rrbracket  = 0, \label{Jacobi}
\end{align}
where $ A_{\hat{a}}, B_{\hat{a}}, C_{\hat{a}} $ are homogeneous elements of $\mathfrak{g}_{\hat{a}}$ and 
\begin{equation}
	\hat{a} + \hat{b} = [a_1+b_1 ,a_2+b_2] \in \Z2, \qquad 
	\hat{a} \cdot \hat{b} = a_1 b_1 + a_2 b_2 \in \mathbb{Z}_2,
\end{equation}
then $\mathfrak{g}$ is referred to as a $\Z2$-graded color Lie superalgebra.

It is clear from the definition that the graded Lie brackets are realized by commutators and anticommutators as follows 
\begin{equation}
	\llbracket A_{\hat{a}}, B_{\hat{b}} \rrbracket
	= 
	\begin{cases}
		[A_{\hat{a}}, B_{\hat{b}}], & \DP{a}{b} = 0,
		\\[10pt]
		\{ A_{\hat{a}}, B_{\hat{b}} \}, & \DP{a}{b} = 1.
	\end{cases}
\end{equation}
If $\llbracket A, B \rrbracket = 0, $ we say that $A$ and $ B$ are $\Z2$-\textit{graded commutative}. 
It is also observed from the definition that $ \g$ has $\mathbb{Z}_2$-grading, too:
\begin{equation}
	\g = \g_0 \oplus \g_1, \qquad \g_0 := \g_{[00]} \oplus \g_{[11]}, \ 
	\g_1 := \g_{[10]} \oplus \g_{[01]}.
\end{equation}

For a given Lie (super)algebra, one may consider its $\Z2$-graded extension which  is generally not unique.  Many inequivalent $\Z2$-graded extensions of the Lie superalgebra $\osp(m|n)$ have been discussed so far \cite{rw2,GrJa,StoVDJ,StoVDJ3,StoVDJosp,AAD,RY}. 
The $\Z2$-graded extension of $\osp(2|2)$ considered in the present work is 16-dimensional and its basis, their $\Z2$-gradings  and their scaling dimensions (eigenvalue of the grading operator $ G:= \frac{1}{2}K_0$) are summarized in the table below:
\begin{equation}
	\begin{array}{c|cccc}
		& [00]   & [10]         & [01]         & [11]
		\\[5pt] \hline
		+1           & K^+    &              &              & L^+
		\\[5pt] 
		+\frac{1}{2} &        & F_{10}^+,~G_{10}^+ & F_{01}^+,~G_{01}^+ &
		\\[5pt] 
		0            & K^0,~R &              &              & L^0,~S
		\\[5pt] 
		-\frac{1}{2} &        & F_{10}^-,~G_{10}^- & F_{01}^-,~G_{01}^- &
		\\[5pt] \
		-1           & K^-    &              &              & L^-
	\end{array}
\end{equation}
We keep using $\g$ to denote this $\Z2$-graded extension of $\osp(2|2).$ 
The non-vanishing defining relations of $\g$ are given, in terms of (anti)commutators, as follows. 

\noindent
$-$ $\g_0$-$\g_0$ sector:
\begin{alignat}{4}
	[K^0, K^{\pm}]     & = \pm 2 K^{\pm},            & \quad
	[K^+, K^-]         & = K^0,                      & \quad
	[L^0, L^{\pm}]     & = \pm 2 K^{\pm},            & \quad
	[L^+, L^-]         & = K^0,
	\nn
	\\[5pt]
	[K^0, L^{\pm}]     & = \pm 2 L^{\pm},            &
	[K^{\pm}, L^{\mp}] & = \pm L^0,                  &
	[L^0, K^{\pm}]     & = \pm 2 L^{\pm}. \label{BB}
\end{alignat}
\

\medskip\noindent
$-$ $\g_0$-$\g_1$ sector:
\begin{alignat}{4}
	[ K^0 , F^{\pm}_{10} ] &= \pm F^{\pm}_{10}, & \quad
	[ K^0 , G^{\pm}_{10} ] &= \pm G^{\pm}_{10}, & \quad
	[ K^0 , F^{\pm}_{01} ] &= \pm F^{\pm}_{01}, & \quad
	[ K^0 , G^{\pm}_{01} ] &= \pm G^{\pm}_{01}, 
	\notag \\
	[ K^{\pm}, F_{10}^{\mp} ] &= -F_{10}^{\pm}, & [ K^{\pm}, G_{10}^{\mp}] &=-G_{10}^{\pm}, & 
	[ K^{\pm}, F_{01}^{\mp} ] &= -F_{01}^{\pm}, & [ K^{\pm}, G_{01}^{\mp}] &=-G_{01}^{\pm},
	\notag\\
	[ R   , F^{\pm}_{10} ] &=   G^{\pm}_{10}, &
	[ R   , G^{\pm}_{10} ] &=   F^{\pm}_{10}, &
	[ R   , F^{\pm}_{01} ] &=   G^{\pm}_{01}, &
	[ R   , G^{\pm}_{01} ] &=   F^{\pm}_{01},
	\notag \\
	\{ L^0 , F^{\pm}_{10} \} &= \pm F^{\pm}_{01}, &
	\{ L^0 , G^{\pm}_{10} \} &= \pm G^{\pm}_{01}, &
	\{ L^0 , F^{\pm}_{01} \} &= \pm F^{\pm}_{10}, &
	\{ L^0 , G^{\pm}_{01} \} &= \pm G^{\pm}_{10}, 
	\notag\\
	\{ L^{\pm}, F_{10}^{\mp} \} &=-F_{01}^{\pm}, & 
	\{ L^{\pm}, G_{10}^{\mp}\} &=-G_{01}^{\pm}, &
	\{ L^{\pm}, F_{01}^{\mp} \} &=-F_{10}^{\pm}, & 
	\{ L^{\pm}, G_{01}^{\mp}\} &=-G_{10}^{\pm},
	\notag\\
	\{ S   , F^{\pm}_{10} \} &=   G^{\pm}_{01}, &
	\{ S   , G^{\pm}_{10} \} &=   F^{\pm}_{01}, &
	\{ S   , F^{\pm}_{01} \} &=   G^{\pm}_{10}, &
	\{ S   , G^{\pm}_{01} \} &=   F^{\pm}_{10}.
\end{alignat}

\medskip\noindent
$-$ $\g_1$-$\g_1$ sector:
\begin{alignat}{4}
	\{ F_{10}^{\pm}, F_{10}^{\pm} \} &= \pm 2K^{\pm}, & \quad \{ G_{10}^{\pm}, G_{10}^{\pm} \} &= \mp 2K^{\pm}, & \quad
	\{ F_{10}^+, F_{10}^- \} &= K^0, & \quad \{ G_{10}^+, G_{10}^- \} &= -K^0,
	\notag \\
	\{F_{10}^{\pm}, G_{10}^{\mp}\} &= \pm R,
	\notag \\
	\{ F_{01}^{\pm}, F_{01}^{\pm} \} &= \mp 2K^{\pm}, & \quad \{ G_{01}^{\pm}, G_{01}^{\pm} \} &= \pm 2K^{\pm}, & \quad
	\{ F_{01}^+, F_{01}^-\} &=-K^0, & \{ G_{01}^+, G_{01}^- \} &= K^0,
	\notag \\
	\{F_{01}^{\pm}, G_{01}^{\mp}\} &= \mp R,
	\notag \\
	[F_{10}^{\pm}, F_{01}^{\pm}] &= \mp 2 L^{\pm}, & [F_{10}^{\pm}, F_{01}^{\mp}] &= -L^0, &  
	[G_{10}^{\pm}, G_{01}^{\pm}] &= \pm 2 L^{\pm}, & [G_{10}^{\pm}, G_{01}^{\mp}] &= L^0,
	\notag \\
	[F_{10}^{\pm}, G_{01}^{\mp}] &= \mp S, & [F_{01}^{\pm}, G_{10}^{\mp}] &= \pm S.
\end{alignat}
It is easy to see that the algebra $\g$ admits  the triangular decomposition;
\begin{align}
	\g &= \g_+ \oplus \h \oplus \g_-,
	\\[3pt]
	\g_+ &= \{  K^+, \ F_{10}^+, \ G_{10}^+, \ F_{01}^+, \ G_{01}^+, \ L^+ \},
	\nn \\[3pt]
	\h &= \{  K^0, \ R, \ L^0, \ S \},
	\nn \\[3pt]
	\g_- &= \{  K^-, \ F_{10}^-, \ G_{10}^-, \ F_{01}^-, \ G_{01}^-,  \ L^- \} \nn
\end{align}
where $\h$ is the commutative subalgebra of $\g$.  
One may also see that the following map $\tau : \g \to \g$ is an outer automorphism for $\epsilon = \pm 1$ (without loss of generality one can set $\epsilon =1$):
\begin{alignat}{4}
	\tau(K^0) &= K^0, &\qquad  \tau(K^{\pm}) &= K^{\pm}, &\qquad  \tau(R) &= -R, &\qquad 
	\tau(L^0) &= \varepsilon L^0,
	\notag\\
	\tau(L^{\pm}) &= \varepsilon L^{\pm}, &  \tau(S) &= -\varepsilon S, 
	&
	\tau(F_{10}^{\pm}) &= F_{10}^{\pm}, & \tau(G_{10}^{\pm}) &= -G_{10}^{\pm},
	\notag \\
	\tau(F_{01}^{\pm}) &= \varepsilon F_{01}^{\pm}, & \tau(G_{01}^{\pm}) &= -\varepsilon G_{01}^{\pm}.
\end{alignat}
The existence of this outer automorphism allows to introduce the twisted affine $\mathbb{Z}_2^2$-graded ${\widehat{osp(2|2)}}$ color superalgebra following the same steps presented in \cite{ToppanZhang}.

The $\Z2$-graded color Lie superalgebra $\g$ is realized by the (complexified) quaternion and the standard ($\mathbb{Z}_2$-graded) $\osp(2|2)$. 
We denote the basis of $\osp(2|2)$ by
\begin{equation}
	[0] \quad H, \ E^{\pm}, \ R^0 \qquad \qquad [1] \quad F^{\pm}, \ G^{\pm}.
\end{equation} 
It is subject to the non-vanishing relations
\begin{alignat}{3}
	[H, E^{\pm}]           & = \pm 2 E^{\pm},      & \qquad [E^+, E^-]      & = H,         & \qquad
	[R^0, H]               & = [R^0, E^{\pm}] = 0,
	\notag \\
	[H, F^{\pm}] & = \pm F^{\pm},& \qquad [H,G^{\pm}] & = \pm G^{\pm},  & \qquad [R^0, F^{\pm}] & = G^{\pm},
	\notag \\
	[R^0, G^{\pm}]         & = F^{\pm},       & [E^{\pm}, F^{\mp}]     & = -F^{\pm},     & [E^{\pm}, G^{\mp}]    & =-G^{\pm},
	\notag \\
	\{ F^{\pm}, F^{\pm}\}  & = \pm 2 E^{\pm}, & \{ G^{\pm}, G^{\pm}\}  & =\mp 2 E^{\pm}, & \{F^{\pm}, G^{\mp}\}  & =\pm R^0,
	\notag \\
	\{ F^{\pm}, F^{\mp} \} & = H,             & \{ G^{\pm}, G^{\mp} \} & = -H.
\end{alignat}

Now we introduce the $ 4 \times 4 $ complexified quaternionic matrices
\begin{align}
	M_0 &:= \mathbb{I}_2 \otimes \mathbb{I}_2, \quad M_1 := \mathbb{I}_2 \otimes \sigma_1, \quad M_2 := \sigma_1 \otimes \sigma_2, \quad M_3 := \sigma_1 \otimes \sigma_3
,\label{Matrix-M}
\end{align}
which are defined by the $2\times 2$ Identity and the Pauli matrices
\begin{equation}
	\mathbb{I}_2 = \begin{pmatrix} 
		1 & 0 \\ 0 & 1
	\end{pmatrix}, 
	\qquad
	\sigma_1 = 
	\begin{pmatrix}
		0 & 1 \\ 1 & 0
	\end{pmatrix},
	\qquad
	\sigma_2 = 
	\begin{pmatrix}
		0 & -i \\ i & 0
	\end{pmatrix},
	\qquad
	\sigma_3 = 
	\begin{pmatrix}
		1 & 0 \\ 0 & -1
	\end{pmatrix}.
\end{equation}
They satisfy for $i,j=1,2,3$ the relations (the totally antisymmetric structure constant $\epsilon_{ijk}$ is normalized so that $\epsilon_{123}=1$):
\begin{equation}
	M_i M_j = \delta_{ij} M_0+ i \epsilon_{ijk} M_k.
\end{equation}
We identify the $\Z2$-grading of the matrices as follows:
\begin{equation}
	[00]  \; : \; M_0; \qquad  [10] \; : \; M_1 ; \qquad [01] \; : \; M_2; \qquad [11] \; : \; M_3.
\end{equation}
Then, $\g$ is realized by the matrices $M_k$ and $ \osp(2|2)$ generators as follows (Cf. \cite{aizt,aiktt,afitt}):
\begin{alignat}{3}
	K^0       & = M_0 \otimes H,        & \qquad K^{\pm} & = M_0 \otimes E_{\pm},  & \qquad R = M_0 \otimes R^0,
	\notag                                                                                                       \\
	L^0       & = M_3 \otimes H,        & \qquad L^{\pm} & = M_3 \otimes E_{\pm},  & \qquad S = M_3 \otimes R^0,
	\notag                                                                                                       \\
	F_{10}^{\pm} & = M_1 \otimes F^{\pm},  & G_{10}^ {\pm}      & = M_1 \otimes G^{\pm},
	\notag                                                                                                       \\
	F_{01}^{\pm} & = iM_2 \otimes F^{\pm}, & G_{01}^{\pm}      & = iM_2 \otimes G^{\pm}. 
	\label{MatrixReal}
\end{alignat}

\section{$\mathcal{N}=2$ extended $\Z2$-super-Liouville equation by Polyakov's soldering} \label{SEC:Equatoion}
\setcounter{equation}{0} 

We mimic the standard procedure of soldering for deriving \cite{polyak} (see also \cite{afitt}) the $\Z2$-graded extension of the $\mathcal{N}=2$ super-Liouville equation \cite{IvaKri}. We introduce the $\Z2$-graded color Lie group $\Z2$-$Osp(2|2)$ generated by the algebra $\g$ defined in \S\ref{SEC:alg}. A group element of $\Z2$-$Osp(2|2)$ is parametrized by 
\begin{align}
	g &= \exp(k_{00}^+ K^+ + l_{11}^+ L^+) \exp(f_{10}^+ F_{10}^+  + f_{01}^+ F_{01}^+) \exp(g_{10}^+ G_{10}^+ + g_{01}^+ G_{01}^+)
	\notag \\
	& \times
	\exp(k_{00} K^0 + l_{11} L^0) \exp(r_{00} R + s_{11}S) 
	\notag \\
	& \times
	\exp(f_{10}^- F_{10}^- +  f_{01}^- F_{01}^-) \exp(g_{10}^- G_{10}^- + g_{01}^- G_{01}^-)
	\exp(k_{00}^- K^- + l_{11}^- L^-) 
	\label{groupelem}
\end{align}
where the group parameters ($k_{00}^{\pm}, l_{11}^{\pm}$ etc.) are also $\Z2$-commutative and their grading is indicated by the suffix. 
In particular, $ f_{10}^{\pm}, g_{10}^{\pm}, f_{01}^{\pm} $ and $ g_{01}^{\pm}$ are nilpotent. 
Throughout this article, the suffices 00, 10, 01, 11 indicate the $\Z2$-grading of the associated quantities.  
We assume that the group parameters are superfields which are functions of the [00] and [10]-graded $\Z2$-commutative variables
\begin{equation}
	[00] \quad x_{\pm},  \qquad [10] \quad \theta_{\pm}. \label{10coordinates}
\end{equation}
Alternatively, one could assume that the parameters are superfields on [00] and [01]-graded variables. 
It is obvious that both assumptions lead to the same equation, so we consider only the case of [00] and [10]-graded variables. 

We introduce the [10]-graded covariant derivatives
\begin{equation}
	D_{\pm} = \partial_{\theta_{\pm}} + i \theta_{\pm} \partial_{\pm}, \qquad \partial_{\pm} := \partial_{x_{\pm}}
\end{equation}
which satisfy
\begin{align}
	\{ D_{\pm}, D_{\pm} \} & = 2i  \partial_{\pm}, \qquad \{ D_+, D_- \}  = 0. 
\end{align}
We introduce also the holomorphic and antiholomorphic WZNW-currents which are defined in terms of the group element \eqref{groupelem}:
\begin{equation}
	J(x_+,\theta_+) := D g\cdot g^{-1}, \qquad \bar{J}(x_-,\theta_-) := g^{-1} \bar{D} g. 
	\label{DEF:current} 
\end{equation}
By definition, the currents $J(x_+,\theta_+), \bar{J}(x_-,\theta_-)$ are [10]-graded and take values in $\g.$  
Employing the matrix presentation \eqref{MatrixReal}, one may rearrange the components of the currents in terms of the  $\osp(2|2)$ generators.  
First, the group element \eqref{groupelem} is given by
\begin{align}
		g&=
		e^{e^+ \otimes E^+}e^{f^+ \otimes F^+}e^{g^+ \otimes G^+}
		e^{h \otimes H}e^{r^0 \otimes R^0}
		e^{g^- \otimes G^-}e^{f^- \otimes F^-}e^{e^- \otimes E^-},
\end{align}
where the matrix valued fields $ e^{\pm}, f^{\pm}, g^{\pm}, h $ and $ r^0 $ are defined by
\begin{alignat}{3}
	e^\pm&= k_{00}^\pm M_0 + l_{11}^\pm M_3,\quad&
	h&=k_{00} M_0 +l_{11} M_3,\quad& r^0&=r_{00} M_0 +s_{11} M_3
	\nonumber\\
	f^\pm&= f_{10}^\pm M_1 +if_{01}^\pm M_2,\quad&
	g^\pm&= g_{10}^\pm M_1 +ig_{01}^\pm M_2  \label{z2fieldsdef}
\end{alignat}
and $ f^{\pm}, g^{\pm}$ are nilpotend due to the nilpotency of $ f_{10}^{\pm}, g_{10}^{\pm}, f_{01}^{\pm} $ and $ g_{01}^{\pm}.$ 
We also see that $ f^+, f^-, g^+$ and $ g^-$ are anticommute. 
It follows immediately that the currents \eqref{DEF:current} takes the following form:
\begin{align}
	J(x_+,\theta_+)
	&=J^{++} \otimes E^+ +J^+_1 \otimes  F^+ +J^+_2 \otimes  G^+
	+J^0 \otimes  H +J^u\otimes  R^0
	\notag \\
	& + J^-_2 \otimes G^- +J^-_1 \otimes F^-  +J^{--} \otimes E^-,
	\notag \\
	\bar J(x_-,\theta_-)
	&=\bar J^{++} \otimes  E^+ +\bar J^+_1 \otimes  F^+ +\bar J^+_2 \otimes  G^+
	+\bar J^0 \otimes  H +\bar J^u\otimes  R^0
	\notag \\
	&+\bar J^-_2 \otimes  G^- +\bar J^-_1\otimes  F^-  +\bar J^{--} \otimes  E^-
	.
\end{align}
where the components of $ J $ and $\bar{J}$ are given by
\begin{align}
	J^{++} &=
	D_+e^+ +D_+f^+\cdot f^+ -D_+g^+\cdot g^+ -2D_+h\cdot e^+ -2D_+r^0\cdot g^+f^+
	,\nonumber\\&
	-2J^-_1f^+e^+ +2J^-_2g^+e^+ -J^{--}(e^+)^2
	,\nonumber\\[5pt]
	J^+_1&=
	D_+f^+ -D_+h\cdot f^+ -D_+r^0 \cdot g^+
	-J^-_1e^+ +J^-_2g^+f^+
	,\nonumber\\[5pt]
	J^+_2&=
	D_+g^+ -D_+h\cdot g^+ -D_+r^0 \cdot f^+
	-J^-_2e^+ +J^-_1g^+f^+ 
	,\nonumber\\[5pt]
	J^0&=
	D_+h +J^-_1f^+ -J^-_2g^+ +J^{--}e^+
	,\nonumber\\[5pt]
	J^u&=
	D_+r^0 +J^-_2f^+ -J^-_1g^+ -J^{--}g^+f^+
	,\nonumber\\[5pt]
	J^-_2&=
	e^{-h}\big[ D_+g^- \cosh r^0 +D_+f^- \sinh r^0 \big]
	+J^{--} g^+
	,\nonumber\\[5pt]
	J^-_1&=
	e^{-h}\big[ D_+f^- \cosh r^0 +D_+g^- \sinh r^0 \big]
	+J^{--} f^+
	,\nonumber\\[5pt]
	J^{--}&=
	e^{-2h}\big[ D_+g^-\cdot g^-
	-D_+f^-\cdot f^-
	+D_+e^- \big]
	, \label{CurrentOspbasis}
\end{align}
and
\begin{align}
	\bar J^{++} &= 
	e^{-2h}\big[ D_-e^+ -f^+D_-f^+ +g^+D_-g^+ \big]
	,\nonumber\\[5pt]
	\bar J^+_1&= 
	e^{-h}\big[ D_-f^+ \cosh r^0 -D_-g^+ \sinh r^0 \big]
	-\bar J^{++} f^-
	,\nonumber\\[5pt]
	\bar J^+_2&=
	e^{-h}\big[ D_-g^+ \cosh r^0 -D_-f^+ \sinh r^0 \big]
	-\bar J^{++} g^-
	,\nonumber\\[5pt]
	\bar J^0&=
	D_-h -\bar J^+_1 f^- +\bar J^+_2 g^- +\bar J^{++} e^-
	,\nonumber\\[5pt]
	\bar J^u&=
	D_-r^0 +\bar J^+_2 f^- -\bar J^+_1 g^- +\bar J^{++} g^-f^-
	,\nonumber\\[5pt]
	\bar J^-_2&=
	-D_-h\cdot g^- +D_-r^0 \cdot f^- +D_-g^- +\bar J^+_2 e^- -\bar J^+_1 g^-f^- 
	,\nonumber\\[5pt]
	\bar J^-_1&=
	-D_-h\cdot f^- +D_-r^0 \cdot g^- +D_-f^- +\bar J^+_1 e^- -\bar J^+_2 g^-f^- 
	,\nonumber\\[5pt]
	\bar J^{--}&=
	-2D_-h\cdot e^- +2D_-r^0\cdot g^-f^- -g^-D_-g^- +f^-D_-f^- +D_-e^-
	\nonumber\\&
	+2\bar J^+_1f^-e^- -2\bar J^+_2g^-e^- -\bar J^{++}(e^-)^2
	.
\end{align}

According to \cite{polyak,sab1,sab2} we impose Hamiltonian constraints on the currents. 
Taking into account the grading and the matrix nature of the components of $ J, \bar{J}$, the appropriate constraints will be (Cf.\cite{afitt})
\begin{alignat}{2}
	J^-_1&=M_1,\qquad 
	J^{--}&=J^-_2=J^0=J^u=J^+_1=0 ,
	\nonumber\\
	\bar J^+_1&=M_1,\qquad 
	\bar J^{++}&=\bar J^+_2=\bar J^0=\bar J^u=\bar J^-_1=0 . 
	\label{Reductionconstraint2}
\end{alignat}
The constraints $ J_1^- = M_1 $ and $ J_2^- = 0$, together with $ J^{--} =0$,  give the relations
\begin{equation}
		D_+f^- = M_1e^h\cosh r^0, \qquad D_+g^- = -M_1e^h\sinh r^0, \label{DFm}
\end{equation}
The constraints $\bar{J}^0 = \bar{J}^u = 0, $ together with $\bar{J}_1^+ = M_1, \bar{J}_2^+ = \bar{J}^{++} = 0,$ give the relations
\begin{equation}
	D_+ D_- h = M_1 D_+ f^-, \qquad 	D_+D_- r^0 = M_1 D_+g^-. \label{DDh}
\end{equation}
Substituting \eqref{DFm} into \eqref{DDh}, we obtain the following system of equations:
\begin{alignat}{2}
	D_+D_- h&= e^h \cosh r^0, \qquad&
	D_+D_- r^0&= -e^h \sinh r^0.
\end{alignat}
Recalling the definition of $ h, r^0 $ in \eqref{z2fieldsdef}, these equations gives a system of equations for $\Z2$-commutative superfields: 
\begin{align}
	D_+ D_- k_{00}
	&=e^{k_{00}}\left( \cosh r_{00} \cosh l_{11} \cosh s_{11} +\sinh r_{00} \sinh l_{11} \sinh s_{11} \right)
	,\nonumber\\
	D_+ D_- l_{11}
	&=e^{k_{00}}\left( \cosh r_{00} \sinh l_{11} \cosh s_{11} +\sinh r_{00} \cosh l_{11} \sinh s_{11} \right)
	,\nonumber\\
	D_+ D_- r_{00}
	&=-e^{k_{00}^0}\left( \sinh r_{00} \cosh l_{11} \cosh s_{11} +\cosh r_{00} \sinh l_{11} \sinh s_{11} \right)
	,\nonumber\\
	D_+ D_- s_{11}
	&=-e^{k_{00}^0}\left( \cosh r_{00} \cosh l_{11} \sinh s_{11} +\sinh r_{00} \sinh l_{11} \cosh s\right)
	.\label{sold_Liouville}
\end{align}
This system is the $\mathcal{N}=2 $ extended $\Z2$-super-Liouville equations. 
We observe that, by setting $ r_{00} = s_{11} = 0, $ these equations reduce to the $\Z2$-super-Liouville equations associated with the $\Z2$-$\osp(1|2)$ discussed in \cite{afitt}. 

As we show in $\S$\ref{sec:zc-eq}, the system \eqref{sold_Liouville} admits a zero-curvature formulation in terms of $\mathbb Z_2^2$-graded Lax operators. This establishes its integrability and leads to further $\mathbb Z_2^2$-graded integrable reductions.

\section{ $\Z2$-graded extension of $\mathcal{N}= 2$ super-Virasoro algebras} \label{SEC:CurrentAlg}
\setcounter{equation}{0}

In this Section we consider the current algebra associated with the currents given in \eqref{DEF:current}.  
This is a $\Z2$-graded extension of the works by Polyakov \cite{polyak} and Sabra \cite{sab1,sab2} along the line in \cite{aiktt} and \cite{afitt}. 
This analysis is carried out within the framework of classical mechanics, i.e., making use of the Poisson brackets. 
Nevertheless, as in the case of  \cite{aiktt,afitt}, we observe the existence of a central term in the Poisson Lie algebra (an example, see \cite{Anomaly}, of a classical anomaly). 

Only the holomorphic sector will be considered here, as the treatment of the antiholomorphic sector proceeds in a similar fashion. 
The transformations of the component currents defined in \eqref{CurrentOspbasis} are induced from the left action of the group element: 
\begin{align}
	g &\to g' = g_{\epsilon} g,
	\notag \\
	g_{\epsilon}&=
	e^{\epsilon^{++} E^+}e^{\epsilon^+_1 F^+}e^{\epsilon^+_2 G^+}
	e^{\epsilon^0 H}e^{\epsilon^u R^0}
	e^{\epsilon^-_2 G^-}e^{\epsilon^-_1 F^-}e^{\epsilon^{--} E^-}
\end{align}
where $ \epsilon$'s are holomorphic functions and $ \epsilon^{\pm}_1, \epsilon^{\pm}_2 $ are nilpotent and anticommute. 
Considering the infinitesimal transformation of $g$, one may obtain
\begin{align}
	\delta J^{++} &=
	-2\epsilon^{++}J^0 +2\epsilon^+_1J^+_1 -2\epsilon^+_2J^+_2 +2\epsilon^0J^{++} +D_+\epsilon^{++}
	,\\
	\delta J^+_1 &=
	-\epsilon^{++}J^-_1 +\epsilon^+_1J^0 +\epsilon^+_2J^u +\epsilon^0J^+_1 +\epsilon^u J^+_2 -\epsilon^-_1J^{++} +D_+\epsilon^+_1
	,\\
	\delta J^+_2 &=
	-\epsilon^{++}J^-_2 +\epsilon^+_1J^u +\epsilon^+_2J^0 +\epsilon^0J^+_2 +\epsilon^u J^+_1 -\epsilon^-_2J^{++} +D_+\epsilon^+_2
	,\\
	\delta J^0 &=
	\epsilon^{++}J^{--} +\epsilon^+_1J^-_1 -\epsilon^+_2J^-_2 -\epsilon^-_2J^+_2 +\epsilon^-_1 J^+_1 -\epsilon^{--}J^{++} +D_+\epsilon^0
	,\\
	\delta J^u &=
	\epsilon^+_1J^-_2 -\epsilon^+_2J^-_1 +\epsilon^-_2J^+_1 -\epsilon^-_1J^+_2 +D_+\epsilon^u
	,\\
	\delta J^-_2 &=
	-\epsilon^+_2J^{--} -\epsilon^0J^-_2 +\epsilon^u J^-_1 -\epsilon^-_2J^0 +\epsilon^-_1J^u -\epsilon^{--}J^+_2 +D_+\epsilon^-_2 
	,\\
	\delta J^-_1 &=
	-\epsilon^+_1J^{--} -\epsilon^0J^-_1 +\epsilon^u J^-_2 +\epsilon^-_2J^u -\epsilon^-_1J^0 -\epsilon^{--}J^+_1 +D_+\epsilon^-_1 
	,\\
	\delta J^{--} &=
	-2\epsilon^0J^{--} +2\epsilon^-_2J^-_2 -2\epsilon^-_1J^-_1 +2\epsilon^{--}J^0 +D\epsilon^{--}
	. \label{CTcomponents}
\end{align}
By the constraints given in \eqref{Reductionconstraint2}, 
it turns out that the infinitesimal transformations preserving \eqref{Reductionconstraint2} have only two independent parameters $\epsilon^{--}$ and $\epsilon_2^-$:
\begin{align}
	\epsilon^-_1&= 
	\frac12 D_+\epsilon^{--}M_1
	,\nonumber\\
	\epsilon^0&=
	\frac{i}{2}\partial_+\epsilon^{--}
	,\nonumber\\
	\epsilon^u&=
	\epsilon^{--}J^+_2M_1 -D_+\epsilon^-_2M_1
	,\nonumber\\
	\epsilon^+_2&=
	\frac12 (D_+\epsilon^{--})J^+_2 +\epsilon^{--}D_+J^+_2-i\partial_+\epsilon^-_2
	,\nonumber\\
	\epsilon^+_1&=
	\epsilon^-_2J^+_2M_1 +\epsilon^{--}J^{++}M_1 -\frac{i}{2}\partial_+D_+\epsilon^{--}M_1
	,\nonumber\\
	\epsilon^{++}&=
	\epsilon^{--}J^+_2J^+_2 -\epsilon^-_2D_+J^+_2 
	+\frac12 (D_+\epsilon^{--})J^{++} +\epsilon^{--}D_+J^{++} +\frac12 \partial_+^2\epsilon^{--}
	.
\end{align}
So that the current transformation \eqref{CTcomponents} reduces to the followings:
\begin{align}
	\delta J^{++} &=
	\frac{3i}{2}(\partial_+ \epsilon^{--})J^{++} +i\epsilon^{--}\partial_+ J^{++} +\frac12 (D_+\epsilon^{--})D_+J^{++}
	\nonumber\\
	&+2i(\partial_+ \epsilon^-_2)J^+_2 +i\epsilon^-_2 \partial_+ J^+_2 -(D_+\epsilon^-_2)D_+J^+_2 
	+\frac12 \partial_+^2D_+\epsilon^{--}
	,\\
	\delta J^+_2 &=
	i(\partial_+\epsilon^{--})J^+_2 +i\epsilon^{--}\partial_+J^+_2 +\frac12 (D_+\epsilon^{--})D_+J^+_2 -\epsilon^-_2 J^{++} +i\partial_+D_+\epsilon^-_2
	.
\end{align}

Recall that $J_{++}$ and $J^+_2$ consist of $\mathbb{Z}_2$-graded superfields and matrices $M_i$, and hence admit an expansion in powers of $\theta_+$.
\begin{equation}
	J^{++}(x_+,\theta_+)=\Gamma(x_+) +\theta_+ T(x_+), 
	\qquad
	J^+_2(x_+,\theta_+) =U(x_+) +\theta_+ \Lambda(x_+).
\end{equation}
One may also expand the transformation parameters
\begin{equation}
	\epsilon^{--}(x_+,\theta_+)=\varepsilon_1(x_+)+\theta_+\varepsilon_2(x_+),
	\qquad 
	\epsilon^{-}_2(x_+,\theta_+)=\tau_1(x_+)+\theta_+\tau_2(x_+),
\end{equation}
By assigning the scaling dimension $ [x_+]=-1,$ twe see that $[\theta_+]= -\frac{1}{2}$ and 
the scaling dimensions of the remaining quantities are determined as follows:
\begin{alignat}{4}
	[\Gamma ]&=\frac32, & \quad [T]&=2, &\quad [U]&=1, &\quad  [\Lambda ]&=\frac32,
	\nonumber\\
	[\varepsilon_1]&=-1,& [\varepsilon_2]&=-\frac12,& [\tau_1]&=-\frac12,& [\tau_2]&=0. 
	\label{SDforCurrents}
\end{alignat}
The transformations of the components are readily obtained
\begin{align}
	\delta T&=
	2i\varepsilon_1' T +i\varepsilon_1 T' 
	+\frac{3i}{2}\varepsilon_2' \Gamma  +\frac{i}{2}\varepsilon_2 \Gamma '
	-3i\tau_1'\Lambda  -i\tau_1\Lambda '
	+2i\tau_2'U +\frac{i}{2}\varepsilon_1'''
	,\nonumber\\
	\delta \Gamma &=
	\frac{3i}{2}\varepsilon_1'\Gamma  +i\varepsilon_1\Gamma ' +\frac12 \varepsilon_2T
	+2i\tau_1'U +i\tau_1U' -\tau_2\Lambda  
	+\frac12 \varepsilon_2''
	,\nonumber\\
	\delta \Lambda &= \frac{3i}{2}\varepsilon_1' \Lambda  +i\varepsilon_1\Lambda '
	+i\varepsilon_2'U  +\frac{i}{2}\varepsilon_2U' 
	+\tau_1T -\tau_2\Gamma  
	-\tau_1''
	,\nonumber\\
	\delta U&=i\varepsilon_1' U +i\varepsilon_1 U' +\frac12 \varepsilon_2\Lambda  -\tau_1\Gamma  +i\tau_2' 
	,\label{transf}
\end{align}
where the prime stands for the derivative with respect to $x_+:$ $\epsilon' = \partial_{x_+}\epsilon. $

From the component currents $\Gamma, T, U $ and $\Lambda$, one may extract the $\Z2$-graded currents by
\begin{alignat}{2}
	\Gamma(x_+) &= \Gamma_{10}(x_+)M_0 +\Gamma_{01}(x_+)M_3
	,\qquad&
	T(x_+) &= T_{00}(x_+)M_0 +T_{11}(x_+)M_3
	,\nonumber\\
	U(x_+) &= U_{00}(x_+)M_1 +iU_{11}(x_+)M_2
	,\qquad&
	\Lambda(x_+) &= \Lambda_{10}(x_+)M_1 +i\Lambda_{01}(x_+)M_2.
\end{alignat}
The $\Z2$-graded parameters are introduced in a similar manner:
\begin{alignat}{2}
	\varepsilon_1(x_+) &= \varepsilon_{00}(x_+)M_0 +\varepsilon_{11}(x_+)M_3
	,&\qquad 
	\varepsilon_2(x_+) &= \varepsilon_{10}(x_+)M_0 +\varepsilon_{01}(x_+)M_3
	,\nonumber\\
	\tau_1(x_+) &= \tau_{10}(x_+)M_1 +i\tau_{01}(x_+) M_2
	,&
	\tau_2(x_+) &= \tau_{00}(x_+)M_1 +i\tau_{11}(x_+)M_2
	.
\end{alignat}
The transformations of the $\Z2$-graded currents are deduced from \eqref{transf}:
\begin{align}
	\delta T_{00}&=
	2i\varepsilon_{00}' T_{00} +i\varepsilon_{00} T_{00}' 
	+\frac{3i}{2}\varepsilon_{10}' \Gamma_{10}  +\frac{i}{2}\varepsilon_{10} \Gamma_{10} '
	-3i\tau_{10}'\Lambda_{10}  -i\tau_{10}\Lambda_{10} '
	+2i\tau_{00}'U_{00}
	\nonumber\\
	&+2i\varepsilon_{11}' T_{11} +i\varepsilon_{11} T_{11}' 
	-\frac{3i}{2}\varepsilon_{01}' \Gamma_{01}  -\frac{i}{2}\varepsilon_{01} \Gamma_{01} '
	+3i\tau_{01}'\Lambda_{01}  +i\tau_{01}\Lambda_{01} '
	+2i\tau_{11}'U_{11}
	+\frac{i}{2}\varepsilon_{00}'''
	,\nonumber\\
	\delta T_{11}&=
	2i\varepsilon_{00}' T_{11} +i\varepsilon_{00} T_{11}' 
	+\frac{3i}{2}\varepsilon_{10}' \Gamma_{01}  +\frac{i}{2}\varepsilon_{10} \Gamma_{01} '
	-3i\tau_{10}'\Lambda_{01}  -i\tau_{10}\Lambda_{01} '
	+2i\tau_{00}'U_{11}
	\nonumber\\
	&+2i\varepsilon_{11}' T_{00} +i\varepsilon_{11} T_{00}' 
	-\frac{3i}{2}\varepsilon_{01}' \Gamma_{10}  -\frac{i}{2}\varepsilon_{01} \Gamma_{10} '
	+3i\tau_{01}'\Lambda_{10}  +i\tau_{01}\Lambda_{10} '
	+2i\tau_{11}'U_{00}
	+\frac{i}{2}\varepsilon_{11}'''
	,\nonumber\\
	\delta \Gamma_{10} &=
	\frac{3i}{2}\varepsilon_{00}'\Gamma_{10}  +i\varepsilon_{00}\Gamma_{10} ' +\frac12 \varepsilon_{10}T_{00}
	+2i\tau_{10}'U_{00} +i\tau_{10}U_{00}' -\tau_{00}\Lambda_{10}  
	\nonumber\\
	&-\frac{3i}{2}\varepsilon_{11}'\Gamma_{01}  -i\varepsilon_{11}\Gamma_{01} ' +\frac12 \varepsilon_{01}T_{11}
	+2i\tau_{01}'U_{11} +i\tau_{01}U_{11}' +\tau_{11}\Lambda_{01}  
	+\frac12 \varepsilon_{10}''
	,\nonumber\\
	\delta \Gamma_{01} &=
	\frac{3i}{2}\varepsilon_{00}'\Gamma_{01}  +i\varepsilon_{00}\Gamma_{01} ' +\frac12 \varepsilon_{10}T_{11}
	+2i\tau_{10}'U_{11} +i\tau_{10}U_{11}' -\tau_{00}\Lambda_{01}  
	\nonumber\\
	&-\frac{3i}{2}\varepsilon_{11}'\Gamma_{10}  -i\varepsilon_{11}\Gamma_{10} ' +\frac12 \varepsilon_{01}T_{00}
	+2i\tau_{01}'U_{00} +i\tau_{01}U_{00}' +\tau_{11}\Lambda_{10}  
	+\frac12 \varepsilon_{01}''
	,\label{Z22transf1}
\end{align}
and
\begin{align}
	\delta \Lambda_{10} &= 
	\frac{3i}{2}\varepsilon_{00}' \Lambda_{10}  +i\varepsilon_{00}\Lambda_{10} '
	+i\varepsilon_{10}'U_{00}  +\frac{i}{2}\varepsilon_{10}U_{00}' 
	+\tau_{10}T_{00} -\tau_{00}\Gamma_{10}  
	\nonumber\\
	&-\frac{3i}{2}\varepsilon_{11}' \Lambda_{01}  -i\varepsilon_{11}\Lambda_{01} '
	+i\varepsilon_{01}'U_{11}  +\frac{i}{2}\varepsilon_{01}U_{11}' 
	+\tau_{01}T_{11} +\tau_{11}\Gamma_{01}  
	-\tau_{10}''
	,\nonumber\\
	\delta \Lambda_{01} &= 
	\frac{3i}{2}\varepsilon_{00}' \Lambda_{01}  +i\varepsilon_{00}\Lambda_{01} '
	+i\varepsilon_{10}'U_{11}  +\frac{i}{2}\varepsilon_{10}U_{11}' 
	+\tau_{10}T_{11} -\tau_{00}\Gamma_{01}  
	\nonumber\\
	&-\frac{3i}{2}\varepsilon_{11}' \Lambda_{10}  -i\varepsilon_{11}\Lambda_{10} '
	+i\varepsilon_{01}'U_{00}  +\frac{i}{2}\varepsilon_{01}U_{00}'
	+\tau_{01}T_{00} +\tau_{11}\Gamma_{10}  
	-\tau_{01}''
	,\nonumber\\
	\delta U_{00}&=
	i\varepsilon_{00}' U_{00} +i\varepsilon_{00} U_{00}' +\frac12 \varepsilon_{10}\Lambda_{10}  -\tau_{10}\Gamma_{10}
	\nonumber\\
	&+i\varepsilon_{11}' U_{11} +i\varepsilon_{11} U_{11}' -\frac12 \varepsilon_{01}\Lambda_{01}  +\tau_{01}\Gamma_{01}  +i\tau_{00}' 
	,\nonumber\\
	\delta U_{11}&=
	i\varepsilon_{00}' U_{11} +i\varepsilon_{00} U_{11}' +\frac12 \varepsilon_{10}\Lambda_{01}  -\tau_{10}\Gamma_{01}
	\nonumber\\
	&+i\varepsilon_{11}' U_{00} +i\varepsilon_{11} U_{00}' -\frac12 \varepsilon_{01}\Lambda_{10}  +\tau_{01}\Gamma_{10}  +i\tau_{11}' 
	.\label{Z22transf2}
\end{align}

It follows from these results that the $\Z2$-graded currents $ T_a,\Gamma_b ,\Lambda_b$ and $ U_a $ $(a\in\{00,11\},~b\in\{10,01\})$ can be identified with a $\Z2$-graded generalization of the $\mathcal{N}=2$ Virasoro algebra. 
This can be seen from the fact that the infinitesimal transformations \eqref{Z22transf1} and \eqref{Z22transf2} are reproduced by the Poisson brackets:
\begin{align}
	\delta A(x_+) &=\int dy_+~ \{iK(y_+), A(x_+)\},
	\nonumber\\
	K(y_+)&= 
	\varepsilon_{00}T_{00} +\varepsilon_{11}T_{11} 
	+\varepsilon_{10}\Gamma_{10} -\varepsilon_{01}\Gamma_{01}
	+2(-\tau_{10}\Lambda_{10} +\tau_{01}\Lambda_{01}  
	+\tau_{00}U_{00} +\tau_{11}U_{11}),
	\label{deltaPoi}
\end{align}
where $A(x_+)\in \{T_a,\Gamma_b ,\Lambda_b ,U_a\},\ a\in\{00,11\},~b\in\{10,01\}$. 
The $\Z2$-graded Poisson bracket is defined by the relations
\begin{align}
	\left\{u_{\hat{a}},\, u_{\hat{b}}\right\}
	& =-(-1)^{\hat{a}\cdot\hat{b}}\left\{u_{\hat{b}},\, u_{\hat{a}}\right\},\label{PB_def}
	\notag\\
	\left\{u_{\hat{a}}u_{\hat{b}},\, u_{\hat{c}}\right\}
	& =u_{\hat{a}}\left\{u_{\hat{b}},\, u_{\hat{c}}\right\}
	+(-1)^{\hat{b}\cdot\hat{c}}\left\{u_{\hat{a}},\, u_{\hat{c}}\right\}u_{\hat{b}},
	\notag\\
	\left\{u_{\hat{a}},\, \left\{u_{\hat{b}},\, u_{\hat{c}}\right\}\right\}
	& =\left\{\left\{u_{\hat{a}},\, u_{\hat{b}}\right\},\, u_{\hat{c}}\right\}
	+(-1)^{\hat{a}\cdot\hat{b}}\left\{u_{\hat{b}},\, \left\{u_{\hat{a}},\, u_{\hat{c}}\right\}\right\},
\end{align}
where $\hat{a},~\hat{b},~\hat{c}\in \Z2.$ 
We note that the Poisson brackets $\{ u_{10}, u_{11} \}, \{ u_{01}, u_{11} \}, \{ u_{10}, u_{10} \} $ and $\{ u_{01}, u_{01} \}$ are symmetric, but all others are antisymmetric. 
The Poisson brackets between the $\Z2$-graded currents can be be derived in a manner similar to that used in \cite{afitt}. They are given by
\begin{align}
	\{T_{00},T_{00}\}&=
	-2T_{00}\delta'(y_+-x_+) -T_{00}'\delta(y_+-x_+) -\tfrac12\delta'''(y_+-x_+),
	\nonumber\\
	\{T_{00},T_{11}\}&=
	-2T_{11}\delta'(y_+-x_+) -T_{11}'\delta(y_+-x_+),
	\nonumber\\
	\{T_{00},\Gamma_{10}\}&=
	-\tfrac32\Gamma_{10}\delta'(y_+-x_+) -\tfrac12\Gamma_{10}'\delta(y_+-x_+),
	\nonumber\\
	\{T_{00},\Gamma_{01}\}&=
	-\tfrac32\Gamma_{01}\delta'(y_+-x_+) -\tfrac12\Gamma_{01}'\delta(y_+-x_+),
	\nonumber\\
	\{T_{00},\Lambda_{10}\}&=
	-\tfrac32\Lambda_{10}\delta'(y_+-x_+) -\tfrac12\Lambda_{10}'\delta(y_+-x_+),
	\nonumber\\
	\{T_{00},\Lambda_{01}\}&=
	-\tfrac32\Lambda_{01}\delta'(y_+-x_+) -\tfrac12\Lambda_{01}'\delta(y_+-x_+),
	\nonumber\\
	\{T_{00},U_{00}\}&=
	-U_{00}\delta'(y_+-x_+),
	\nonumber\\
	\{T_{00},U_{11}\}&=
	-U_{11}\delta'(y_+-x_+),
	\nonumber\\[3pt]
	\{T_{11},T_{11}\}&=
	-2T_{00}\delta'(y_+-x_+) -T_{00}'\delta(y_+-x_+) -\tfrac12\delta'''(y_+-x_+),
	\nonumber\\
	\{T_{11},\Gamma_{10}\}&=
	\tfrac32\Gamma_{01}\delta'(y_+-x_+) +\tfrac12\Gamma_{01}'\delta(y_+-x_+),
	\nonumber\\
	\{T_{11},\Gamma_{01}\}&=
	\tfrac32\Gamma_{10}\delta'(y_+-x_+) +\tfrac12\Gamma_{10}'\delta(y_+-x_+),
	\nonumber\\
	\{T_{11},\Lambda_{10}\}&=
	\tfrac32\Lambda_{01}\delta'(y_+-x_+) +\tfrac12\Lambda_{01}'\delta(y_+-x_+),
	\nonumber\\
	\{T_{11},\Lambda_{01}\}&=
	\tfrac32\Lambda_{10}\delta'(y_+-x_+) +\tfrac12\Lambda_{10}'\delta(y_+-x_+),
	\nonumber\\
	\{T_{11},U_{00}\}&=
	-U_{11}\delta'(y_+-x_+),
	\nonumber\\
	\{T_{11},U_{11}\}&=
	-U_{00}\delta'(y_+-x_+),
	\nonumber\\[3pt]
	\{\Gamma_{10},\Gamma_{10}\}&=
	-\tfrac{i}{2}T_{00}\delta(y_+-x_+) -\tfrac{i}{2}\delta''(y_+-x_+),
	\nonumber\\
	\{\Gamma_{10},\Gamma_{01}\}&=
	-\tfrac{i}{2}T_{11}\delta(y_+-x_+),
	\nonumber\\
	\{\Gamma_{10},\Lambda_{10}\}&=
	-U_{00}\delta'(y_+-x_+) -\tfrac12U_{00}'\delta(y_+-x_+),
	\nonumber\\
	\{\Gamma_{10},\Lambda_{01}\}&=
	-U_{11}\delta'(y_+-x_+) -\tfrac12U_{11}'\delta(y_+-x_+),
	\nonumber\\
	\{\Gamma_{10},U_{00}\}&=
	-\tfrac{i}{2}\Lambda_{10}\delta(y_+-x_+),
	\nonumber\\
	\{\Gamma_{10},U_{11}\}&=
	-\tfrac{i}{2}\Lambda_{01}\delta(y_+-x_+),
	\nonumber\\[3pt]
	\{\Gamma_{01},\Gamma_{01}\}&=
	\tfrac{i}{2}T_{00}\delta(y_+-x_+) +\tfrac{i}{2}\delta''(y_+-x_+),
	\nonumber\\
	\{\Gamma_{01},\Lambda_{10}\}&=
	U_{11}\delta'(y_+-x_+) +\tfrac12U_{11}'\delta(y_+-x_+),
	\nonumber\\
	\{\Gamma_{01},\Lambda_{01}\}&=
	U_{00}\delta'(y_+-x_+) +\tfrac12U_{00}'\delta(y_+-x_+),
	\nonumber\\
	\{\Gamma_{01},U_{00}\}&=
	-\tfrac{i}{2}\Lambda_{01}\delta(y_+-x_+),
	\nonumber\\
	\{\Gamma_{01},U_{11}\}&=
	-\tfrac{i}{2}\Lambda_{10}\delta(y_+-x_+),
	\nonumber\\[3pt]
	\{\Lambda_{10},\Lambda_{10}\}&=
	\tfrac{i}{2}T_{00}\delta(y_+-x_+) -\tfrac{i}{2}\delta''(y_+-x_+),
	\nonumber\\
	\{\Lambda_{10},\Lambda_{01}\}&=
	\tfrac{i}{2}T_{11}\delta(y_+-x_+),
	\nonumber\\
	\{\Lambda_{10},U_{00}\}&=
	-\tfrac{i}{2}\Gamma_{10}\delta(y_+-x_+),
	\nonumber\\
	\{\Lambda_{10},U_{11}\}&=
	-\tfrac{i}{2}\Gamma_{01}\delta(y_+-x_+),
     \nonumber\\[3pt]   
	\{\Lambda_{01},\Lambda_{01}\}&=
	-\tfrac{i}{2}T_{00}\delta(y_+-x_+) +\tfrac{i}{2}\delta''(y_+-x_+),
	\nonumber\\
	\{\Lambda_{01},U_{00}\}&=
	-\tfrac{i}{2}\Gamma_{01}\delta(y_+-x_+),
	\nonumber\\
	\{\Lambda_{01},U_{11}\}&=
	-\tfrac{i}{2}\Gamma_{10}\delta(y_+-x_+),
	\nonumber\\[3pt]
	\{U_{00},U_{00}\}&=
	-\tfrac12\delta'(y_+-x_+),
	\nonumber\\
	\{U_{00},U_{11}\}&=
	0,
	\nonumber\\
	\{U_{11},U_{11}\}&=
	-\tfrac12\delta'(y_+-x_+), \label{CurrentPoi}
\end{align}
where the prime denotes the derivative with respect to $y_+$,  
and our convention for the delta function is
\begin{align}
	\delta(x-a) &=  \sum_{n \in \mathbb{Z}} e^{in(x-a)},
	\\
	\frac{1}{2\pi}\oint dx \delta(x) &= \frac{1}{2\pi} \int_0^{2\pi} dx \delta(x) = 1.
\end{align}
These relations define a $\Z2$-graded extension of the $\mathcal{N} =2$ super-Virasoro algebra. 
Note that the ordinary $\mathcal{N} = 2$ 	super-Virasoro algebra is recovered by either $ T_{00}, U_{00}, \Gamma_{10}, \Lambda_{10}$ or $ T_{00}, U_{00}, \Gamma_{01}, \Lambda_{01}$ subalgebras.

By expanding the $\Z2$-graded currents into modes, we obtain from \eqref{CurrentPoi} the infinite dimensional $\Z2$-graded super-Virasoro algebra equipped with a Poisson-Lie structure. 
As is well known in string theory, fermionic sectors are subject to either periodic (Ramond sector) or antiperiodic (Neveu-Schwarz sector) boundary conditions \cite{Ramond,NS}. 
Assuming, as usual, that the [00]-graded currents are periodic, consistency with the current algebra \eqref{CurrentPoi} imposes specific (anti)periodicity conditions on the remaining graded sectors. 
One should note that even the [11]-graded currents could satisfy either periodic or antiperiodic boundary condition.  
Therefore, three possible alternatives are admissible in association with 
$
[11]/[10]/[01] 
$
graded sectors:
\begin{center}
	(i)  R/R/R \qquad (ii) R/NS/NS \qquad (iii) NS/NS/R $\equiv$ NS/R/NS 
\end{center}
where R and NS stand for the Ramond and Neveu-Schwarz sector, respectively. 
We remark that these three admissible boundary conditions already appeared in the string model induced from 1D $\Z2$-graded supersymmetry \cite{aizt}.

Under these periodicity conditions, the $\Z2$-graded currents may be expanded as
\begin{align}
	T_{00}(x)
	& = \sum_{n\in\mathbb Z} T^{00}_n e^{inx},
	&
	T_{11}(x)
	& = \sum_{p\in\mathbb Z} T^{11}_p e^{ipx},
	\nn
	\\
	\Gamma_{10}(x)
	& = \sum_{r\in\mathbb Z} \Gamma^{10}_r e^{irx},
	&
	\Gamma_{01}(x)
	& = \sum_{s\in\mathbb Z} \Gamma^{01}_s e^{isx},
	\nn
	\\
	\Lambda_{10}(x)
	& = \sum_{r\in\mathbb Z} \Lambda^{10}_r e^{irx},
	&
	\Lambda_{01}(x)
	& = \sum_{s\in\mathbb Z} \Lambda^{01}_s e^{isx},
	\nn\\
	U_{00}(x)
	& = \sum_{n\in\mathbb Z} U^{00}_n e^{inx},
	&
	V_{11}(x)
	& = \sum_{p\in\mathbb Z} U^{11}_p e^{ipx},
\end{align}
where $ n \in \mathbb{Z}$ is the index for [00]-sector (Ramond sector) and $ p, r, s$ are the  indices for [11], [10], [01]-sectors, respectively. The indices $p, r, s$ take their values in $\mathbb{Z}$ if they are in the Ramond sector or in $  \mathbb{Z}+\frac{1}{2}$ if in the  Neveu-Schwarz sector.
Then we obtain the relations of the $\Z2$-graded extension of $\mathcal{N}=2$   super-Virasoro algebra:
\begin{alignat}{2}
	\{T^{00}_{n},T^{00}_{n'}\} &= i(n'-n)T^{00}_{n+n'} + \tfrac{i}{2}n^{3}\delta_{n+n',0}, \qquad&
	\{T^{00}_{n},T^{11}_{p}\} &= i(p-n)\,T^{11}_{n+p},\nonumber\\
	\{T^{00}_{n},\Gamma^{10}_{r}\} &= i\!\left(r-\tfrac{n}{2}\right)\Gamma^{10}_{n+r}, \qquad&
	\{T^{00}_{n},\Gamma^{01}_{s}\} &= i\!\left(s-\tfrac{n}{2}\right)\Gamma^{01}_{n+s},\nonumber\\
	\{T^{00}_{n},\Lambda^{10}_{r}\} &= i\!\left(r-\tfrac{n}{2}\right)\Lambda^{10}_{n+r}, \qquad&
	\{T^{00}_{n},\Lambda^{01}_{s}\} &= i\!\left(s-\tfrac{n}{2}\right)\Lambda^{01}_{n+s},\nonumber\\
	\{T^{00}_{n},U^{00}_{n'}\} &= i\,n'\,U^{00}_{n+n'}, \qquad&
	\{T^{00}_{n},U^{11}_{p}\} &= i\,p\,U^{11}_{n+p},\nonumber\\[4pt]
	\{T^{11}_{p},T^{11}_{p'}\} &= i(p'-p)T^{00}_{p+p'} + \tfrac{i}{2}p^{3}\delta_{p+p',0}, \qquad&
	\{T^{11}_{p},\Gamma^{10}_{r}\} &= i\!\left(\tfrac{p}{2}-r\right)\Gamma^{01}_{p+r},\nonumber\\
	\{T^{11}_{p},\Gamma^{01}_{s}\} &= i\!\left(\tfrac{p}{2}-s\right)\Gamma^{10}_{p+s}, \qquad&
	\{T^{11}_{p},\Lambda^{10}_{r}\} &= i\!\left(\tfrac{p}{2}-r\right)\Lambda^{01}_{p+r},\nonumber\\
	\{T^{11}_{p},\Lambda^{01}_{s}\} &= i\!\left(\tfrac{p}{2}-s\right)\Lambda^{10}_{p+s}, \qquad&
	\{T^{11}_{p},U^{00}_{n}\} &= i\,n\,U^{11}_{p+n},\nonumber\\
	\{T^{11}_{p},U^{11}_{p'}\} &= i\,p'\,U^{00}_{p+p'},\nonumber\\[4pt]
	\{\Gamma^{10}_{r},\Gamma^{10}_{r'}\} &= -\tfrac{i}{2}\,T^{00}_{r+r'} + \tfrac{i}{2}\,r^{2}\delta_{r+r',0}, \qquad&
	\{\Gamma^{10}_{r},\Gamma^{01}_{s}\} &= -\tfrac{i}{2}\,T^{11}_{r+s},\nonumber\\
	\{\Gamma^{10}_{r},\Lambda^{10}_{r'}\} &= \tfrac{i}{2}(r'-r)\,U^{00}_{r+r'}, \qquad&
	\{\Gamma^{10}_{r},\Lambda^{01}_{s}\} &= \tfrac{i}{2}(s-r)\,U^{11}_{r+s},\nonumber\\
	\{\Gamma^{10}_{r},U^{00}_{n}\} &= -\tfrac{i}{2}\,\Lambda^{10}_{r+n}, \qquad&
	\{\Gamma^{10}_{r},U^{11}_{p}\} &= -\tfrac{i}{2}\,\Lambda^{01}_{r+p},\nonumber\\[4pt]
	\{\Gamma^{01}_{s},\Gamma^{01}_{s'}\} &= \tfrac{i}{2}\,T^{00}_{s+s'} - \tfrac{i}{2}\,s^{2}\delta_{s+s',0}, \qquad&
	\{\Gamma^{01}_{s},\Lambda^{10}_{r}\} &= \tfrac{i}{2}(s-r)\,U^{11}_{s+r},\nonumber\\
	\{\Gamma^{01}_{s},\Lambda^{01}_{s'}\} &= \tfrac{i}{2}(s-s')\,U^{00}_{s+s'}, \qquad&
	\{\Gamma^{01}_{s},U^{00}_{n}\} &= -\tfrac{i}{2}\,\Lambda^{01}_{s+n},\nonumber\\
	\{\Gamma^{01}_{s},U^{11}_{p}\} &= -\tfrac{i}{2}\,\Lambda^{10}_{s+p},\nonumber\\[4pt]
	\{\Lambda^{10}_{r},\Lambda^{10}_{r'}\} &= \tfrac{i}{2}\,T^{00}_{r+r'} + \tfrac{i}{2}\,r^{2}\delta_{r+r',0}, \qquad&
	\{\Lambda^{10}_{r},\Lambda^{01}_{s}\} &= \tfrac{i}{2}\,T^{11}_{r+s},\nonumber\\
	\{\Lambda^{10}_{r},U^{00}_{n}\} &= -\tfrac{i}{2}\,\Gamma^{10}_{r+n}, \qquad&
	\{\Lambda^{10}_{r},U^{11}_{p}\} &= -\tfrac{i}{2}\,\Gamma^{01}_{r+p},\nonumber\\[4pt]
	\{\Lambda^{01}_{s},\Lambda^{01}_{s'}\} &= -\tfrac{i}{2}\,T^{00}_{s+s'} - \tfrac{i}{2}\,s^{2}\delta_{s+s',0}, \qquad&
	\{\Lambda^{01}_{s},U^{00}_{n}\} &= -\tfrac{i}{2}\,\Gamma^{01}_{s+n},\nonumber\\
	\{\Lambda^{01}_{s},U^{11}_{p}\} &= -\tfrac{i}{2}\,\Gamma^{10}_{s+p}, \nonumber\\
	\{U^{00}_{n},U^{00}_{n'}\} &= -\tfrac{i}{2}\,n\,\delta_{n+n',0}, \qquad&
	\{U^{00}_{n},U^{11}_{p}\} &= 0, \nonumber\\[4pt]
	\{U^{11}_{p},U^{11}_{p'}\} &= -\tfrac{i}{2}\,p\,\delta_{p+p',0}
\end{alignat}
where the indices take the following values according to the boundary conditions. 
(i) all the indices are integer, (ii) $ n, n', p, p' \in \mathbb{Z}$ and $ r, r', s, s' \in \mathbb{Z}+\frac{1}{2},$ (iii) $ n, n', s, s' \in \mathbb{Z}$ and $ p, p', r, r' \in \mathbb{Z}+ \frac{1}{2}. $

We derive the Lie-Poisson algebra with a [00]-graded central extension, in which the central charge takes a fixed value. The most general $\Z2$-graded extension of the super-Virasoro algebra admits two central extensions: one [00]-graded and the other [11]-graded \cite{NAJS}. 
However, the extension admitting a [11]-graded central charge is not obtained within the present framework. 

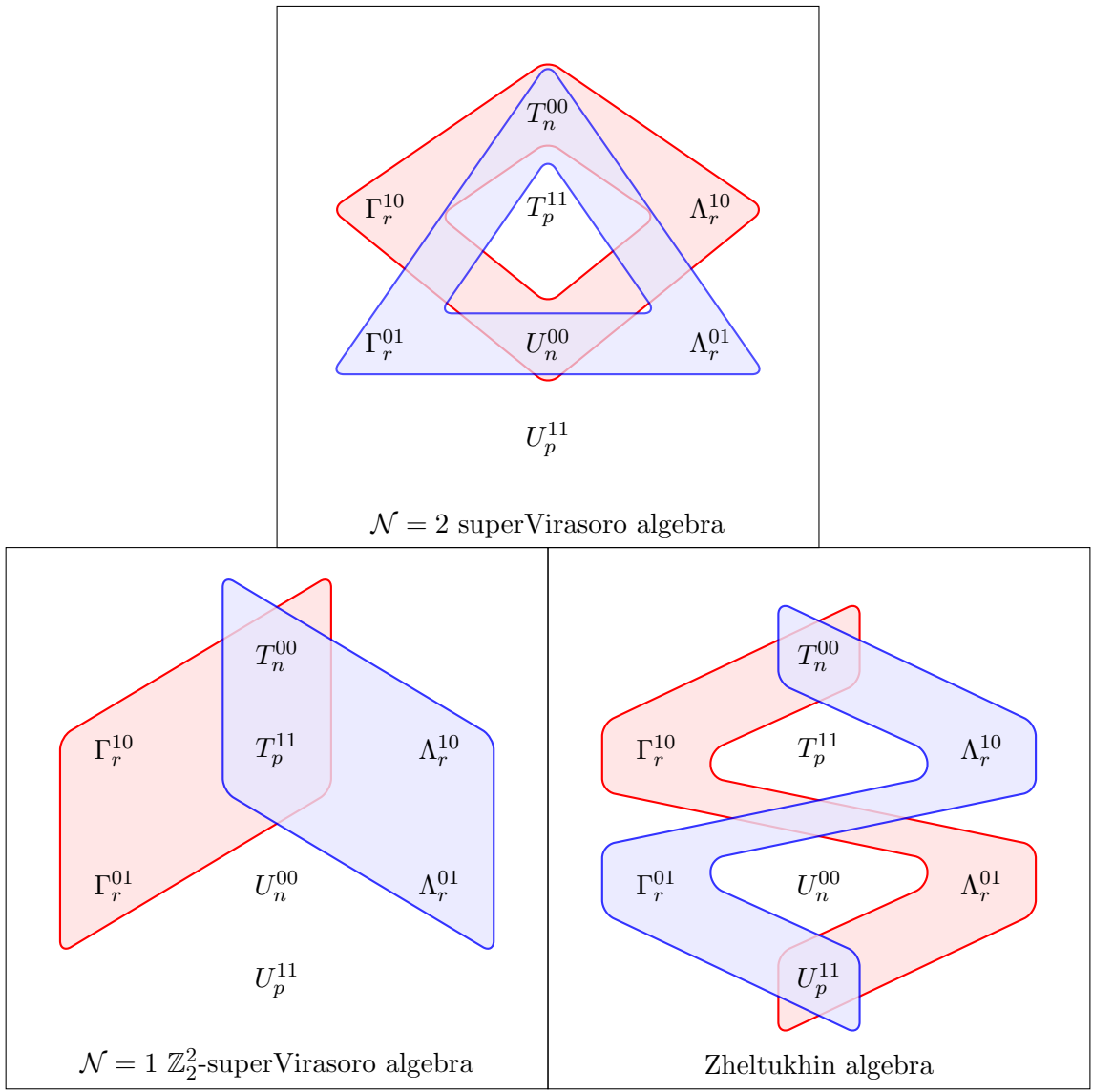
\begin{figure}[h] 
	\begin{center}
		\begin{tikzpicture}[scale=0.75]
			\begin{scope}[shift={(0,0)}]
				\draw (5,5)--(5,-5)--(-5,-5)--(-5,5)--cycle;
				\filldraw[thick, rounded corners=5pt, draw = red,fill=red!10, even odd rule] (0,4)--(4,1.25)--(0,-2)--(-4,1.25)--cycle 
				[scale=0.5,shift={(0,1)}] (0,4)--(4,1.25)--(0,-2)--(-4,1.25)--cycle;
				\filldraw[thick, rounded corners=5pt, draw=blue, fill=blue!10, even odd rule,opacity=0.7,shift={(0,-0.05)}] (0,4)--(4,-1.75)--(-4,-1.75)--cycle
				[scale=0.5, shift={(0,0.5)}] (0,4)--(4,-1.75)--(-4,-1.75)--cycle;
				%
				\node at (0,3) {$T^{00}_n$};
				\node at (-3,1.25) {$\Gamma^{10}_r$};
				\node at (0,1.25) {$T^{11}_p$};
				\node at (3,1.25) {$\Lambda^{10}_r$};
				\node at (-3,-1.25) {$\Gamma^{01}_r$};
				\node at (0,-1.25) {$U^{00}_n$};
				\node at (3,-1.25) {$\Lambda^{01}_r$};
				\node at (0,-3) {$U^{11}_p$};
				\node[above] at (0,-5) {$\mathcal N=2$ superVirasoro algebra};
			\end{scope}
			\begin{scope}[shift={(-5,-10)}]
				\draw (5,5)--(5,-5)--(-5,-5)--(-5,5)--cycle;
				\filldraw[thick, rounded corners=5pt, draw=red,fill=red!10,opacity=1] (1,4.5)--(-4,1.5)--(-4,-2.5)--(1,0.5)--cycle;
				\filldraw[thick, rounded corners=5pt, draw=blue,fill=blue!10,opacity=0.8] (-1,4.5)--(4,1.5)--(4,-2.5)--(-1,0.5)--cycle;
				\node at (0,3) {$T^{00}_n$};
				\node at (-3,1.25) {$\Gamma^{10}_r$};
				\node at (0,1.25) {$T^{11}_p$};
				\node at (3,1.25) {$\Lambda^{10}_r$};
				\node at (-3,-1.25) {$\Gamma^{01}_r$};
				\node at (0,-1.25) {$U^{00}_n$};
				\node at (3,-1.25) {$\Lambda^{01}_r$};
				\node at (0,-3) {$U^{11}_p$};
				\node[above] at (0,-5) {$\mathcal N=1$ $\mathbb Z_2^2$-superVirasoro algebra};
			\end{scope}
			\begin{scope}[shift={(5,-10)}]
				\draw (5,5)--(5,-5)--(-5,-5)--(-5,5)--cycle;
				\filldraw[thick, rounded corners=5pt, draw=red, fill=red!10,opacity=1] 
				(0.75,4)--
				(-4,1.75)--(-4,.5)--
				(2,-.75)--(2,-1.25)--
				(-0.75,-2.5)--(-0.75,-4)--
				(4,-1.75)--(4,-.5)--
				(-2,0.75)--(-2,1.25)--
				(0.75,2.5)--
				cycle;
				\filldraw[thick, rounded corners=5pt, draw=blue,fill=blue!10,opacity=0.8] 
				(-0.75,4)--
				(4,1.75)--(4,.5)--
				(-2,-.75)--(-2,-1.25)--
				(0.75,-2.5)--(0.75,-4)--
				(-4,-1.75)--(-4,-.5)--
				(2,0.75)--(2,1.25)--
				(-0.75,2.5)--
				cycle;
				\node at (0,3) {$T^{00}_n$};
				\node at (-3,1.25) {$\Gamma^{10}_r$};
				\node at (0,1.25) {$T^{11}_p$};
				\node at (3,1.25) {$\Lambda^{10}_r$};
				\node at (-3,-1.25) {$\Gamma^{01}_r$};
				\node at (0,-1.25) {$U^{00}_n$};
				\node at (3,-1.25) {$\Lambda^{01}_r$};
				\node at (0,-3) {$U^{11}_p$};
				\node[above] at (0,-5) {Zheltukhin algebra};
			\end{scope}
		\end{tikzpicture}
		\caption{Inequivalent $6$-generator subalgebras of the $\mathcal N=2$ $\mathbb Z_2^2$-superVirasoro}\label{F1}
	\end{center}
\end{figure}

This algebra contains two different types of $\Z2$-graded extensions of the $\mathcal{N} =1$ super-Virasoro algebra as  subalgebras. 
One of them is discussed in \cite{zhe} and is spanned by either $ T^{00}_n, U^{11}_p, \Gamma^{10}_r, \Lambda^{01}_s$ or $ T^{00}_n, U^{11}_p, \Lambda^{10}_r, \Gamma^{01}_s. $ 
The $(00,11,10,01)$-graded generators have the scaling dimension $(2,1,\frac{3}{2},\frac{3}{2}); $ see  \eqref{SDforCurrents} for the scaling dimensions.   
The other one, which is obtained by soldering for $\Z2$-graded $\osp(1|2)$, is  spanned by either $ T^{00}_n, T^{11}_p, \Gamma^{10}_r, \Gamma^{01}_s$ or $ T^{00}_n, T^{11}_p, \Lambda^{10}_r, \Lambda^{01}_s$ and has the scaling dimension $(2,2,\frac{3}{2},\frac{3}{2})$ \cite{afitt}. 
As already pointed out, the ordinary ($\mathbb{Z}_2$-graded) $\mathcal{N}=2$ super-Virasoro algebra is also a subalgebra.  
These structures are summarized in the table below and Figure~\ref{F1}:
\[
\def\arraystretch{1.5}
  \begin{array}{c|c|c}
  	\Z2\text{-superVirasoro \cite{zhe}}  & \multirow{2}{*}{$T^{00}_n, U^{11}_p, \Gamma^{10}_r, \Lambda^{01}_s$} & \multirow{2}{*}{$T^{00}_n, U^{11}_p, \Lambda^{10}_r, \Gamma^{01}_s$}
  	\\
  	(2,1,\frac{3}{2},\frac{3}{2}) & & 
  	\\ \hline 
  	\Z2\text{-superVirasoro \cite{afitt}}  & \multirow{2}{*}{$T^{00}_n, T^{11}_p, \Gamma^{10}_r, \Gamma^{01}_s$} & \multirow{2}{*}{$T^{00}_n, T^{11}_p, \Lambda^{10}_r, \Lambda^{01}_s $} 
  	\\
  	(2,2,\frac{3}{2},\frac{3}{2}) & & 
  	\\ \hline
  	\mathcal{N}=2 ~\text{superVirasoro} & \multirow{2}{*}{$T^{00}_n, U^{00}_n, \Gamma^{10}_r, \Lambda^{10}_r$} & \multirow{2}{*}{$T^{00}_n, U^{00}_n, \Gamma^{01}_s, \Lambda^{01}_s $} 	
  	\\
  	(2,1,\frac{3}{2},\frac{3}{2})&&
  \end{array}
\] 

\section{Zero-curvature formulation for $\Z2$-graded extension of super-Liouville and sinh-Gordon equations}\label{sec:zc-eq}

In this section, we investigate the integrability of the $\mathcal N=2$ extended $\mathbb Z_2^2$-super-Liouville equations derived in $\S$\ref{SEC:Equatoion} via the soldering procedure. 
We show that these equations admit a zero-curvature formulation, which naturally leads to a more general two-parameter family of integrable equations. 
For a particular choice of parameters, this formulation reproduces the $\mathcal N=2$ extended $\mathbb Z_2^2$-super-Liouville equations, while different choices give rise to $\mathbb Z_2^2$-graded super-sinh-Gordon equations. 
We then construct the corresponding B{\"a}cklund transformations for both systems.

\subsection{Lax operators and zero-curvature equation}

Let us consider superfields defined on the superspace with [00]- and [10]-graded coordinates \eqref{10coordinates}. We introduce the $(1,0)$-graded Lax operators
\begin{align}
\mathcal{L}_\pm
=
\mp D_\pm\Phi
+
e^{\pm\mathrm{ad}\,\Phi}
\left(
a\,F_{10}^{\pm}
+
b\,G_{10}^{\mp}
\right),
\label{Lax_op}
\end{align}
where
\begin{align}
\Phi
=
\frac12
\left(
\Phi_{00}^{S}K_0
+\Phi_{00}^{A}R
+\Phi_{11}^{(1)}L_0
+\Phi_{11}^{(2)}S
\right),
\end{align}
with $\Phi_{00}^{S}$ and $\Phi_{00}^{A}$ being [00]-graded superfields, and $\Phi_{11}^{(1)}$ and $\Phi_{11}^{(2)}$ being [11]-graded superfields. Here $a$ and $b$ are [00]-graded constant parameters.

Following the standard formulation of integrable systems, we consider the associated linear system
\begin{align}
(D_\pm-\mathcal L_\pm)T=0,
\end{align}
where $T$ is a group-valued superfield generated by $\mathfrak g$. The compatibility condition of this linear system yields the zero-curvature equation
\begin{align}
D_+\mathcal L_-+D_-\mathcal L_+
=
\{\mathcal L_+,\mathcal L_-\}.
\label{ZC-eq}
\end{align}

Substituting the explicit expressions of $\mathcal L_\pm$ in \eqref{Lax_op} into \eqref{ZC-eq} and comparing the coefficients of each generator of $\mathfrak g$, we obtain the following four equations of motion:
\begin{align}
D_+ D_- \Phi_{00}^S
&=
\left(a^2 e^{\Phi_{00}^S}-b^2 e^{-\Phi_{00}^S}\right)
\cosh \Phi_{00}^{A}
\cosh \Phi_{11}^{(1)}
\cosh \Phi_{11}^{(2)}
\nn\\
&\qquad
+ \left(a^2 e^{\Phi_{00}^S}+b^2 e^{-\Phi_{00}^S}\right)
\sinh \Phi_{00}^{A}
\sinh \Phi_{11}^{(1)}
\sinh \Phi_{11}^{(2)},
\nn\\[4pt]
D_+ D_- \Phi_{11}^{(1)}
&=
\left(a^2 e^{\Phi_{00}^S}+b^2 e^{-\Phi_{00}^S}\right)
\cosh \Phi_{00}^{A}
\sinh \Phi_{11}^{(1)}
\cosh \Phi_{11}^{(2)}
\nn\\
&\qquad
+ \left(a^2 e^{\Phi_{00}^S}-b^2 e^{-\Phi_{00}^S}\right)
\sinh \Phi_{00}^{A}
\cosh \Phi_{11}^{(1)}
\sinh \Phi_{11}^{(2)},
\nn\\[4pt]
D_+ D_- \Phi_{00}^{A}
&=
-\left(a^2 e^{\Phi_{00}^S}+b^2 e^{-\Phi_{00}^S}\right)
\sinh \Phi_{00}^{A}
\cosh \Phi_{11}^{(1)}
\cosh \Phi_{11}^{(2)}
\nn\\
&\qquad
-\left(a^2 e^{\Phi_{00}^S}-b^2 e^{-\Phi_{00}^S}\right)
\cosh \Phi_{00}^{A}
\sinh \Phi_{11}^{(1)}
\sinh \Phi_{11}^{(2)},
\nn\\[4pt]
D_+ D_- \Phi_{11}^{(2)}
&=
-\left(a^2 e^{\Phi_{00}^S}+b^2 e^{-\Phi_{00}^S}\right)
\cosh \Phi_{00}^{A}
\cosh \Phi_{11}^{(1)}
\sinh \Phi_{11}^{(2)}
\nn\\
&\qquad
-\left(a^2 e^{\Phi_{00}^S}-b^2 e^{-\Phi_{00}^S}\right)
\sinh \Phi_{00}^{A}
\sinh \Phi_{11}^{(1)}
\cosh \Phi_{11}^{(2)}.
\label{fulleqs}
\end{align}

In addition to these equations of motion, the remaining components of the zero-curvature equation yield the following constraints:
\begin{align}
ab\,\sinh \Phi_{00}^{A}\cosh \Phi_{11}^{(2)}&=0,
\nn\\
ab\,\cosh \Phi_{00}^{A}\sinh \Phi_{11}^{(2)}&=0.
\label{constraint}
\end{align}

The system \eqref{fulleqs}, together with the constraints \eqref{constraint}, admit several consistent reductions. In particular, by suitable choices of the parameters $(a,b)$ and reductions of the superfields, one obtains Liouville- and sinh-Gordon-type equations. We now discuss these reductions in detail.

\subsubsection{$\mathcal N=2$ extended $\mathbb Z_2^2$-super-Liouville equations}

By setting $a=1$ and $b=0$, the equations of motion \eqref{fulleqs} reduce to
\begin{align}
D_+ D_- \Phi_{00}^S
&=
 e^{\Phi_{00}^S}
\left(
\cosh \Phi_{00}^{A}
\cosh \Phi_{11}^{(1)}
\cosh \Phi_{11}^{(2)}
+
\sinh \Phi_{00}^{A}
\sinh \Phi_{11}^{(1)}
\sinh \Phi_{11}^{(2)}
\right),
\nn\\
D_+ D_- \Phi_{11}^{(1)}
&=
 e^{\Phi_{00}^S}
\left(
\cosh \Phi_{00}^{A}
\sinh \Phi_{11}^{(1)}
\cosh \Phi_{11}^{(2)}
+
\sinh \Phi_{00}^{A}
\cosh \Phi_{11}^{(1)}
\sinh \Phi_{11}^{(2)}
\right),
\nn\\
D_+ D_- \Phi_{00}^{A}
&=
-  e^{\Phi_{00}^S}
\left(
\sinh \Phi_{00}^{A}
\cosh \Phi_{11}^{(1)}
\cosh \Phi_{11}^{(2)}
+
\cosh \Phi_{00}^{A}
\sinh \Phi_{11}^{(1)}
\sinh \Phi_{11}^{(2)}
\right),
\nn\\
D_+ D_- \Phi_{11}^{(2)}
&=
-  e^{\Phi_{00}^S}
\left(
\cosh \Phi_{00}^{A}
\cosh \Phi_{11}^{(1)}
\sinh \Phi_{11}^{(2)}
+
\sinh \Phi_{00}^{A}
\sinh \Phi_{11}^{(1)}
\cosh \Phi_{11}^{(2)}
\right).
\label{N2-Liouville}
\end{align}
This system gives the $\mathcal N=2$ extended $\mathbb Z_2^2$-super-Liouville equations derived in \eqref{sold_Liouville}. Therefore, the zero-curvature formulation provides the integrability for the $\mathcal N=2$ extended $\mathbb Z_2^2$-super-Liouville equations.

Furthermore, imposing the reduction
\begin{align}
\Phi_{00}^{A}=\Phi_{11}^{(2)}=0,
\end{align}
the above system further reduces to
\begin{align}
D_+D_-\Phi_{00}^{S}
&=
e^{\Phi_{00}^{S}}
\cosh\Phi_{11}^{(1)},
\nn\\
D_+D_-\Phi_{11}^{(1)}
&=
e^{\Phi_{00}^{S}}
\sinh\Phi_{11}^{(1)},
\end{align}
which coincides with the $\mathbb Z_2^2$-super-Liouville equations obtained from the $\mathbb Z_2^2$-$\mathfrak{osp}(1|2)$ zero-curvature formulation in \cite{afitt}.

\subsubsection{$\mathbb Z_2^2$-graded super-sinh-Gordon equations}

Next, we consider another reduction of the system \eqref{fulleqs}. Imposing
\begin{align}
\Phi_{00}^{A}=\Phi_{11}^{(2)}=0,
\end{align}
the system reduces to
\begin{align}
D_+D_-\Phi_{00}^{S}
&=
\left(
a^2e^{\Phi_{00}^{S}}
-
b^2e^{-\Phi_{00}^{S}}
\right)
\cosh\Phi_{11}^{(1)},
\nn\\
D_+D_-\Phi_{11}^{(1)}
&=
\left(
a^2e^{\Phi_{00}^{S}}
+
b^2e^{-\Phi_{00}^{S}}
\right)
\sinh\Phi_{11}^{(1)}.
\end{align}

In particular, choosing
\begin{align}
a^2=b^2=\frac12,
\end{align}
we obtain
\begin{align}
D_+D_-\Phi_{00}^{S}
&=
\sinh\Phi_{00}^{S}\cosh\Phi_{11}^{(1)},
\nn\\
D_+D_-\Phi_{11}^{(1)}
&=
\cosh\Phi_{00}^{S}\sinh\Phi_{11}^{(1)}.
\label{Z22-sinhGordon}
\end{align}
This system may be regarded as a natural $\mathbb Z_2^2$-graded extension of the super-sinh-Gordon equations obtained from the $\mathfrak{osp}(2|2)$ zero-curvature formulation in \cite{ivto}. In fact, by setting $\Phi_{11}^{(1)}=0$, the system reduces to the ordinary super-sinh-Gordon equations.

To study the component expression of \eqref{Z22-sinhGordon}, we expand the superfields as
\begin{align}
\Phi_{00}^{S}
&=
\varphi_{00}
+\theta_+\psi_{10}^{-}
+\theta_-\psi_{10}^{+}
+\theta_+\theta_-A_{00},
\nn\\
\Phi_{11}^{(1)}
&=
\varphi_{11}
+\theta_+\psi_{01}^{-}
+\theta_-\psi_{01}^{+}
+\theta_+\theta_-A_{11}.
\end{align}

Substituting these expansions into \eqref{Z22-sinhGordon}, we obtain the corresponding component equations:
\begin{align}
	\partial_{+-} \varphi_{00}
	&=- A_{00}\cosh \varphi_{00} \cosh \varphi_{11} -A_{11}\sinh \varphi_{00} \sinh \varphi_{11}
	\nn\\
	&-\left( 
	\psi_{10}^+\psi_{10}^- -\psi_{01}^+\psi_{01}^-
	\right) \sinh \varphi_{00} \cosh \varphi_{11}
	-\left(
	\psi_{10}^+\psi_{01}^- -\psi_{01}^+\psi_{10}^-
	\right)\cosh \varphi_{00} \sinh \varphi_{11}
	,\nn\\[5pt]
	i\partial_+ \psi_{10}^+
	&= \psi_{10}^- \cosh \varphi_{00} \cosh \varphi_{11} +\psi_{01}^- \sinh \varphi_{00} \sinh \varphi_{11}
	,\nn\\[5pt]
	i\partial_- \psi_{10}^-
	&= -\psi_{10}^+ \cosh \varphi_{00} \cosh \varphi_{11} -\psi_{01}^+ \sinh \varphi_{00} \sinh \varphi_{11}
	,\nn\\[5pt]
	A_{00} 
	&=- \sinh \varphi_{00} \cosh \varphi_{11}
	,
\end{align}
and 
\begin{align}
	\partial_{+-} \varphi_{11}
	&=- A_{00}\sinh \varphi_{00} \sinh \varphi_{11} -A_{11}\cosh \varphi_{00} \cosh \varphi_{11}
	\nn\\
	&-\left( 
	\psi_{10}^+\psi_{10}^- -\psi_{01}^+\psi_{01}^-
	\right) \cosh \varphi_{00} \sinh \varphi_{11}
	-\left(
	\psi_{10}^+\psi_{01}^- -\psi_{01}^+\psi_{10}^-
	\right)\sinh \varphi_{00} \cosh \varphi_{11}
	,\nn\\[5pt]
	i\partial_+ \psi_{01}^+
	&= \psi_{10}^- \sinh \varphi_{00} \sinh \varphi_{11} +\psi_{01}^- \cosh \varphi_{00} \cosh \varphi_{11}
	,\nn\\[5pt]
	i\partial_- \psi_{01}^-
	&= -\psi_{10}^+ \sinh \varphi_{00} \sinh \varphi_{11} -\psi_{01}^+ \cosh \varphi_{00} \cosh \varphi_{11}
	,\nn\\[5pt]
	A_{11} 
	&=- \cosh \varphi_{00} \sinh \varphi_{11}
	.
\end{align}
Eliminating the auxiliary fields $A_{00}$ and $A_{11}$, we finally arrive at the component form of the $\mathbb Z_2^2$-graded super-sinh-Gordon equations:
\begin{align}
	\partial_{+-} \varphi_{00}
	&=\frac12 \sinh 2\varphi_{00} \cosh 2\varphi_{11}
	-\left( 
	\psi_{10}^+\psi_{10}^- -\psi_{01}^+\psi_{01}^-
	\right) \sinh \varphi_{00} \cosh \varphi_{11}
	\nn\\
	&-\left(
	\psi_{10}^+\psi_{01}^- -\psi_{01}^+\psi_{10}^-
	\right)\cosh \varphi_{00} \sinh \varphi_{11}
	,\nn\\[5pt]
	\partial_{+-} \varphi_{11}
	&=\frac12 \cosh 2\varphi_{00} \sinh 2\varphi_{11}
	-\left( 
	\psi_{10}^+\psi_{10}^- -\psi_{01}^+\psi_{01}^-
	\right) \cosh \varphi_{00} \sinh \varphi_{11}
	\nn\\
	&-\left(
	\psi_{10}^+\psi_{01}^- -\psi_{01}^+\psi_{10}^-
	\right)\sinh \varphi_{00} \cosh \varphi_{11}
	,\nn\\[5pt]
	i\partial_+ \psi_{10}^+
	&= \psi_{10}^- \cosh \varphi_{00} \cosh \varphi_{11} +\psi_{01}^- \sinh \varphi_{00} \sinh \varphi_{11}
	,\nn\\[5pt]
	i\partial_+ \psi_{01}^+
	&= \psi_{10}^- \sinh \varphi_{00} \sinh \varphi_{11} +\psi_{01}^- \cosh \varphi_{00} \cosh \varphi_{11}
	,\nn\\[5pt]
	i\partial_- \psi_{10}^-
	&= -\psi_{10}^+ \cosh \varphi_{00} \cosh \varphi_{11} -\psi_{01}^+ \sinh \varphi_{00} \sinh \varphi_{11}
	,\nn\\[5pt]
	i\partial_- \psi_{01}^-
	&= -\psi_{10}^+ \sinh \varphi_{00} \sinh \varphi_{11} -\psi_{01}^+ \cosh \varphi_{00} \cosh \varphi_{11}
	.
\end{align}

Thus, the reduction \eqref{Z22-sinhGordon} gives a new component realization of a $\mathbb Z_2^2$-graded super-sinh-Gordon equations, which is distinct from that obtained in \cite{AiFujiIto}. In particular, by setting
\begin{align}
\varphi_{11}=\psi_{01}^{\pm}=0,
\end{align}
the system further reduces to the ordinary super-sinh-Gordon equations.

\subsection{B{\"a}cklund transformations for the $\Z2$-graded super-Liouville and sinh-Gordon equations}
Here we construct B{\"a}cklund transformations for the $\mathcal N=2$ extended $\Z2$-super-Liouville equations and the $\Z2$-graded super-sinh-Gordon equations. These transformations provide an equivalent description of the equations of motion and play an important role in the study of their integrable structures. We also derive the corresponding conserved currents associated with these B{\"a}cklund transformations.

\subsubsection{$\mathcal N=2$ extended $\Z2$-super-Liouville equations}

Let us introduce the [00]-graded matrix-valued superfields
\begin{align}
\Phi_1^{\pm}
&=
\Phi_{00}^{S}M_0 \pm \Phi_{11}^{(1)}M_3,
\qquad
\Phi_2^{\pm}
=
\Phi_{00}^{A}M_0 \pm \Phi_{11}^{(2)}M_3,
\label{Matrixfield}
\end{align}
where $M_i$'s are defined in \eqref{Matrix-M}. This parametrization combines the component superfields into matrix-valued objects adapted to the $\mathbb Z_2^2$-graded structure. In terms of these variables, the equations take a compact form.

Using these superfields, the $\mathcal N=2$ extended $\Z2$-super-Liouville equations \eqref{N2-Liouville} can be rewritten as
\begin{align}
D_+D_- \Phi_1^{\pm} &= e^{\Phi_1^{\pm}}\cosh\Phi_2^{\pm},
\qquad
D_+D_- \Phi_2^{\pm} = -e^{\Phi_1^{\pm}}\sinh\Phi_2^{\pm}.
\end{align}
These forms closely resemble the ordinary $\mathcal N=2$ super-Liouville equations. The matrix-valued description therefore allows us to construct the corresponding B{\"a}cklund transformation in a way parallel to the conventional $\mathcal N=2$ case.

To formulate the B{\"a}cklund transformation, we further introduce the auxiliary superfields
\begin{alignat}{2}
\tilde \Phi_1^{\pm}
&=
\tilde \Phi_{00}^{S}M_0 \pm \tilde \Phi_{11}^{(1)}M_3,
\qquad&
\tilde \Phi_2^{\pm}
&=
\tilde \Phi_{00}^{A}M_0 \pm \tilde \Phi_{11}^{(2)}M_3,
\notag \\
\Theta^{\pm}
&=
\Theta_{10}M_0 \pm \Theta_{01}M_3,
\qquad&
\Pi^{\pm}
&=
\Pi_{10}M_0  \pm\Pi_{01}M_3, \label{AugSuperF}
\end{alignat}
and define
\begin{alignat}{2}
V_\pm &= \frac12(\Phi_1^+\pm\tilde \Phi_1^+),
\qquad&
W_\pm &= \frac12(\Phi_2^+\pm\tilde \Phi_2^+),
\notag \\
U_{\pm} &= \frac12(\Phi_1^-\pm\tilde \Phi_1^-), 
&
Z_{\pm} &= \frac12(\Phi_2^-\pm\tilde \Phi_2^-).
\end{alignat}
The B{\"a}cklund transformation is then given by
\begin{align}
D_+V_+
&=
\frac{a}{4}
\left(
e^{V_-+W_-}\Theta^+
+
e^{V_- -W_-}\Pi^+
\right),
\notag\\
D_-V_-
&=
\frac14
\left(
e^{V_+ +W_+}\Pi^+
+
e^{V_+ -W_+}\Theta^+
\right),
\notag\\
D_+W_+
&=
-\frac{a}{4}
\left(
e^{V_-+W_-}\Theta^+
-
e^{V_- -W_-}\Pi^+
\right),
\notag\\
D_-W_-
&=
-\frac14
\left(
e^{V_+ +W_+}\Pi^+
-
e^{V_+ -W_+}\Theta^+
\right),
\notag\\
D_+\Theta^+
&=
e^{V_- -W_-},
\qquad
D_-\Theta^+
=
-\frac1a e^{V_+ +W_+},
\notag\\
D_+\Pi^+
&=
e^{V_- +W_-},
\qquad
D_-\Pi^+
=
-\frac1a e^{V_+ -W_+},
\label{Backlund}
\end{align}
and
\begin{align}
	D_+U_+
	&=
	\frac{a}{4}
	\left(
	e^{U_-+Z_-}\Theta^-
	+
	e^{U_- -Z_-}\Pi^-
	\right),
	\notag\\
	D_-U_-
	&=
	\frac14
	\left(
	e^{U_+ +Z_+}\Pi^-
	+
	e^{U_+ -Z_+}\Theta^-
	\right),
	\notag\\
	D_+Z_+
	&=
	-\frac{a}{4}
	\left(
	e^{U_-+Z_-}\Theta^-
	-
	e^{U_- -Z_-}\Pi^-
	\right),
	\notag\\
	D_-Z_-
	&=
	-\frac14
	\left(
	e^{U_+ +Z_+}\Pi^-
	-
	e^{U_+ -Z_+}\Theta^-
	\right),
	\notag\\
	D_+\Theta^-
	&=
	e^{U_- -Z_-},
	\qquad
	D_-\Theta^-
	=
	-\frac1a e^{U_+ +Z_+},
	\notag\\
	D_+\Pi^-
	&=
	e^{U_- +Z_-},
	\qquad
	D_-\Pi^-
	=
	-\frac1a e^{U_+ -Z_+},
	\label{BacklundMinus}
\end{align}
where $a$ is the B{\"a}cklund parameter.

The compatibility conditions of the equations in \eqref{Backlund} and \eqref{BacklundMinus} reproduce the equations of motion for $\Phi_1^{\pm}$ and $\Phi_2^{\pm}$. Moreover, if $\tilde{\Phi}_1^{\pm}$ and $\tilde{\Phi}_2^{\pm}$ satisfy \eqref{Backlund} and \eqref{BacklundMinus}, then they obey the free equations
\begin{equation}
D_+D_-\tilde{\Phi}_1^{\pm}
=
D_+D_-\tilde{\Phi}_2^{\pm}
=
0.
\label{FreeEq}
\end{equation}
Thus, the B{\"a}cklund transformation relates solutions of the interacting system to those of the corresponding free theory.

Besides this transformation, which connects the interacting theory with free superfields, one can also construct an auto-B{\"a}cklund transformation connecting two $\mathcal N=2$ extended $\Z2$-super-Liouville equations:
\begin{align}
D_+V_+
&=
\frac{a}{2}
\Bigl(
\cosh(V_-+W_-)\Theta^+
+
\cosh(V_- -W_-)\Pi^+
\Bigr),
\notag\\
D_-V_-
&=
\frac12
\Bigl(
e^{V_+ +W_+}\Pi^+
+
e^{V_+ -W_+}\Theta^+
\Bigr),
\notag\\
D_+W_+
&=
-\frac{a}{2}
\Bigl(
\cosh(V_-+W_-)\Theta^+
-
\cosh(V_- -W_-)\Pi^+
\Bigr),
\notag\\
D_-W_-
&=
-\frac12
\Bigl(
e^{V_+ +W_+}\Pi^+
-
e^{V_+ -W_+}\Theta^+
\Bigr),
\notag\\
D_+\Theta^+
&=
\sinh(V_- -W_-),
\qquad
D_-\Theta^+
=
-\frac1a e^{V_+ +W_+},
\notag\\
D_+\Pi^+
&=
\sinh(V_-+W_-),
\qquad
D_-\Pi^+
=
-\frac1a e^{V_+ -W_+},
\label{AutoBack}
\end{align}
and
\begin{align}
	D_+U_+
	&=
	\frac{a}{2}
	\Bigl(
	\cosh(U_-+Z_-)\Theta^-
	+
	\cosh(U_- -Z_-)\Pi^-
	\Bigr),
	\notag\\
	D_-U_-
	&=
	\frac12
	\Bigl(
	e^{U_+ +Z_+}\Pi^-
	+
	e^{U_+ -Z_+}\Theta^-
	\Bigr),
	\notag\\
	D_+Z_+
	&=
	-\frac{a}{2}
	\Bigl(
	\cosh(U_-+Z_-)\Theta^-
	-
	\cosh(U_- -Z_-)\Pi^-
	\Bigr),
	\notag\\
	D_-Z_-
	&=
	-\frac12
	\Bigl(
	e^{U_+ +Z_+}\Pi^-
	-
	e^{U_+ -Z_+}\Theta^-
	\Bigr),
	\notag\\
	D_+\Theta^-
	&=
	\sinh(U_- -Z_-),
	\qquad
	D_-\Theta^-
	=
	-\frac1a e^{U_+ +Z_+},
	\notag\\
	D_+\Pi^-
	&=
	\sinh(U_-+Z_-),
	\qquad
	D_-\Pi^-
	=
	-\frac1a e^{U_+ -Z_+}.
	\label{AutoBackMinus}
\end{align}
If $\tilde{\Phi}_1^{\pm}$ and $\tilde{\Phi}_2^{\pm}$ satisfy \eqref{AutoBack} and \eqref{AutoBackMinus}, then they also satisfy the $\mathcal N=2$ extended $\Z2$-super-Liouville equation \eqref{N2-Liouville}. Hence \eqref{AutoBack} and \eqref{AutoBackMinus} map a solution of the equation to another solution of the same equation.

From these B{\"a}cklund systems, one can naturally construct conserved currents. We define the following $\Z2$-graded currents:
\begin{alignat}{2}
	J_{00}^{\pm} &= D_{\mp} (\Theta_{10} + \Pi_{10}), & \qquad 
	J_{11}^{\pm} &= D_{\mp} (\Theta_{01} + \Pi_{01}),
	\notag \\[3pt]
	J_{10}^{\pm} &= D_{\mp} (\Theta_{10} \Pi_{10} - \Theta_{01} \Pi_{01}), & 
	J_{01}^{\pm} &= D_{\mp} (\Theta_{10} \Pi_{01} - \Theta_{01} \Pi_{10}).
\end{alignat}
Then, it can be verified immediately that these currents satisfy the conservation law
\begin{align}
D_+J_{\hat a}^+ +D_-J_{\hat a}^-=0,
\qquad
\hat a\in\mathbb Z_2^2.
\end{align}

\subsubsection{$\Z2$-graded super-sinh-Gordon equations}

We next consider the $\Z2$-graded super-sinh-Gordon equations \eqref{Z22-sinhGordon}. 
Using the matrix-valued superfield $\Phi_1^{\pm}$ introduced in \eqref{Matrixfield}, the equation \eqref{Z22-sinhGordon} can be rewritten in the compact form
\begin{align}
D_+D_-\Phi_1^{\pm}=\sinh\Phi_1^{\pm}.
\end{align}
The corresponding auto-B{\"a}cklund transformation is given by
\begin{alignat}{2}
D_+V_+
&=
a\Theta^+\cosh V_-,
\qquad&
D_-V_-
&=
\Theta^+\cosh V_+,
\nn\\
D_+\Theta^+
&=
\sinh V_-,
\qquad&
D_-\Theta^+
&=
-\frac1a\sinh V_+,
\end{alignat}
and 
\begin{alignat}{2}
	D_+U_+
	&=
	a\Theta^-\cosh U_-,
	\qquad&
	D_-U_-
	&=
	\Theta^-\cosh U_+,
	\nn\\
	D_+\Theta^-
	&=
	\sinh U_-,
	\qquad&
	D_-\Theta^-
	&=
	-\frac1a\sinh U_+,
\end{alignat}
where $ \Theta^{\pm} $ is defined in \eqref{AugSuperF}. 
The compatibility condition of this system reproduces the $\Z2$-graded sinh-Gordon equations. Furthermore, if $\tilde{\Phi}_1^{\pm}$ satisfies this transformation, then they also satisfy \eqref{Z22-sinhGordon}. Therefore, this B{\"a}cklund transformation maps a solution of the $\Z2$-graded sinh-Gordon equation to another solution of the same equation.

As in the Liouville case, this B{\"a}cklund transformations also yield $\Z2$-graded conserved currents. They are defined by
\begin{alignat}{2}
	J^{\pm}_{00} &= D_{\mp} \Theta_{10}, &  J^{\pm}_{11} &= D_{\mp} \Theta_{01}, 
	\notag \\[3pt]
	J^{\pm}_{10} &= \pm (D_{\mp} \Theta_{10} \cdot \Theta_{10} - D_{\mp} \Theta_{01}\cdot \Theta_{01}),
	& \quad 
	J^{\pm}_{01} &= \pm (D_{\mp} \Theta_{10} \cdot \Theta_{01} - D_{\mp} \Theta_{01}\cdot \Theta_{10}).
\end{alignat} 
It is easy to verify that they satisfy the conservation law
\begin{align}
D_+J_{\hat a}^+ +D_-J_{\hat a}^-=0,
\qquad
\hat a\in\mathbb Z_2^2.
\end{align}

\section{Reproduction of the $\mathbb Z_2^2$-graded sine-Gordon equation}
\label{sec:Z22-sineGordon}

In \cite{bruSG}, Bruce introduced a $\Z2$-graded extension of the sine-Gordon equation and showed its integrability by means of a B\"acklund transformation. 
A superfield formulation was also employed to construct the $\Z2$-graded extension.  However, the underlying superspace has a [11]-graded coordinate, implying that the superfield contains infinitely many component fields in its expansion in the nontrivially graded coordinates. 
In this section, we show that Bruce's $\Z2$-graded sine-Gordon equation  has a connection with the $\mathfrak{osp}(2|2)$ zero-curvature formulation by showing that the lowest-order equations of motion can be reproduced from the Lax operators.

We consider the $\Z2$-graded superspace
\[
    (x_+,x_-,\xi,\eta),
\]
where the Grassmann coordinates $\xi$ and $\eta$ have degrees $[10]$
and $[01]$, respectively. 
The corresponding covariant derivatives are defined by
\begin{align}
    D_{10}=\partial_\xi-\frac12\xi\partial_+,
    \qquad
    D_{01}=\partial_\eta-\frac12\eta\partial_- .
\end{align}
They satisfy
\begin{align}
    D_{10}^2=-\frac12\partial_+,\qquad
    D_{01}^2=-\frac12\partial_-,
    \qquad
    [D_{10},D_{01}]=0 .
\end{align}

Let $\Phi_{00}(x_+,x_-,\xi,\eta)$ be a $[00]$-graded superfield and set 
\begin{align}
    \Phi=\frac{i}{4}\Phi_{00}K^0.
\end{align}
We introduce the following Lax operators:
\begin{alignat}{2}
    \mathcal{L}_{10}
    &=
    -D_{10}\Phi
    +N e^{\operatorname{ad}\Phi}
    \left(F_{10}^+ +\alpha_{11}G_{01}^-\right),
    \nn\\[2mm]
    \mathcal{L}_{01}
    &=
    D_{01}\Phi
    +N e^{-\operatorname{ad}\Phi}
    \left(G_{01}^+ +\alpha_{11}F_{10}^-\right)
,
\end{alignat}
where $N$ is a normalization constant and $\alpha_{11}$ is a
$[11]$-graded constant subject to the condition $ \alpha_{11}^2 = 1$.

Following the standard formulation of integrable systems, we consider the associated linear system
\begin{align}
    (D_{10}-\mathcal{L}_{10})T=0,\quad (D_{01}-\mathcal{L}_{01})T=0,
\end{align}
where $T$ is a group-valued superfield generated by $\mathfrak g$. The compatibility condition of this linear system yields the zero-curvature equation
\begin{align}
    D_{10}\mathcal{L}_{01}
    -D_{01}\mathcal{L}_{10}
    =
    [\mathcal{L}_{10},\mathcal{L}_{01}] .
\end{align}

We obtain the equation of motion for $\Phi_{00}$ from the zero-curvature equation:
\begin{align}
    D_{10}D_{01}\Phi_{00}
    +4N^2\alpha_{11}\sin\frac{\Phi_{00}}{2}=0 . \label{BruceEq}
\end{align}
Setting $N^2 = \frac{1}{8}$, this equation takes exactly the same form as Bruce's $\Z2$-graded sine-Gordon equation. 
However, note that $\Phi_{00}$ is not the same superfield as the one in \cite{bruSG}, since $ \Phi_{00} $ does not depend on the [11]-graded coordinate.

We now write the equation \eqref{BruceEq} in terms of its component fields. 
Expanding the $[00]$-graded superfield as
\begin{align}
    \Phi_{00}
    =
    \varphi_{00}
    +\xi\psi_{10}
    +\eta\psi_{01}
    +\xi\eta F_{11},
\end{align}
and substituting this into \eqref{BruceEq}, we find that  $F_{11}$ is an auxiliary field:  $ 2F_{11} = -\alpha_{11}\sin\frac{\varphi_{00}}{2}$. Eliminating $F_{11}$, we obtain the following component equations:
\begin{align}
    \partial_{+-}\varphi_{00}
    -\frac14\sin\varphi_{00}
    -\frac{\alpha_{11}}{2}\psi_{01}\psi_{10}
    \sin\frac{\varphi_{00}}{2}
    &=0,
    \nn\\
    \partial_+\psi_{01}
    +\frac{\alpha_{11}}{2}\psi_{10}
    \cos\frac{\varphi_{00}}{2}
    &=0,
    \nn\\
    \partial_-\psi_{10}
    +\frac{\alpha_{11}}{2}\psi_{01}
    \cos\frac{\varphi_{00}}{2}
    &=0 .
\end{align}
These equations coincide with the lowest-order component equations of Bruce's
$\mathbb Z_2^2$-graded sine-Gordon equation. 
Thus, the present zero-curvature formulation not only yields the
new $\mathbb Z_2^2$-graded Liouville and sinh-Gordon-type systems
discussed in the previous sections, but also reproduces the previously known
$\mathbb Z_2^2$-graded sine-Gordon equation within the same
$\mathfrak{osp}(2|2)$-based integrable framework.

\section{Conclusion}

We have investigated a superconformal affine Toda theory associated with a $\mathbb{Z}_2^2$-graded extension of the twisted affine superalgebra $\widehat{\osp(2|2)^{(2)}}$.  
We have constructed a model by employing the soldering procedure and a Lax operator formulation based on a $\Z2$-graded $\osp(2|2)$ subalgebra. 
The model interpolates between $\Z2$-graded extensions of the super-Liouville and super-sinh-Gordon equations. We have also derived explicit formulas for the Bäcklund transformations associated with these two equations. 
Furthermore, we have derived a $\Z2$-graded extension of the $\mathcal{N}=2$ super-Virasoro algebra as a Poisson–Lie algebra from the WZNW currents used in the soldering procedure. 
The present results include the previous results of \cite{afitt}, based on the $\Z2$-graded $\osp(1|2)$, as a special case. However, the $\Z2$-graded $\mathcal{N} = 2$ super-Virasoro contains as subalgebras not only the $\Z2$-graded $\mathcal{N}=1$ super-Virasoro algebra introduced in \cite{afitt}, but also the another one introduced  many years ago \cite{zhe}. 
The two $\Z2$-graded $\mathcal{N}=1$ super-Virasoro subalgebras differ in the scaling dimensions of their component fields. 
The present framework also provides a Lax operator formulation of the \(\mathbb{Z}_2\)-graded sine-Gordon equation, which was originally formulated in $\Z2$-graded supersymmetric sigma model \cite{brusigma}.

It is natural to extend the present work to higher rank $\Z2$-graded color Lie (super)algebras and to establish a $\Z2$-graded Toda field theory for  general simple algebras of this type. 
It would also be worthwhile to examine $\Z2$-graded extensions of the construction based on two-loop Kac-Moody algebra (also called \textit{toroidal} Lie algebra \cite{MRY}) initiated in \cite{afgz}.
Despite these challenges, color Lie (super)algebras offer an interesting direction for future research. 
It is known that a superfield formulation based on a $\Z2$-graded superspace leads to a theory defined in a higher-dimensional spacetime \cite{aizt,NARITT2D}. Recall that the present work and previous studies in \cite{aiktt,afitt} have not employed the full $\Z2$-graded superspace, but rather a superspace consisting only of bosonic and parafermionic coordinates. Therefore, employing the full $\Z2$-graded superspace may lead to integrable systems defined in higher-dimensional spacetimes. 
Most studies of $\Z2$-graded integrable systems to date have focused on $(1+1)$-dimensional spacetime, with the only exception we are aware of being \cite{yuan2026super}. 
Extending these studies to higher-dimensional spacetimes would broaden the scope of $\Z2$-graded integrable systems and help clarify the role of color Lie (super)algebras in the construction of integrable field theories beyond $(1+1)$ dimensions.

\bigskip
\section*{Acknowledgments}
The authors are grateful to Z. Kuznetsova for fruitful discussions. 
N. A. and R. I. are grateful to CBPF and UFABC, where this work was initiated, for hospitality. 
N. A. is supported by JSPS KAKENHI Grant Number JP23K03217. 
R. I. is supported by JST SPRING, Grant Number JPMJSP2139. 
F. T. is supported by
CNPq (PQ grant 304747/2026-2).


\begin{thebibliography}{0}
\bibitem{aiktt} N. Aizawa, Ren Ito, Z. Kuznetsova, Toshiya Tanaka and F. Toppan,
\textit{Integrable $\mathbb{Z}_2^2$-graded extensions of the Liouville and Sinh-Gordon theories},
J. Phys. A: Math. Theor. \textbf{58}, 055201 (2025).

\bibitem{rw1} V. Rittenberg and D. Wyler, 
\textit{Generalized superalgebras}, Nucl. Phys. {\bf B 139}, 189 (1978).

\bibitem{rw2} V. Rittenberg and D. Wyler, 
\textit{Sequences of $Z_2\otimes Z_2$ graded Lie algebras and superalgebras}, J. Math. Phys. {\bf 19}, 2193 (1978).

\bibitem{afitt} N. Aizawa, I. Fujii, R. Ito, T. Tanaka and F. Toppan, 
\textit{Integrable $\mathbb{Z}_2^2$-graded super-Liouville Equation and Induced $\mathbb{Z}_2^2$-graded super-Virasoro Algebra}, 
J. Phys. A: Math. Theor. \textbf{59}, 235202 (2026).

\bibitem{babo} O. Babelon and L. Bonora, \textit{Conformal affine $sl_2$ Toda field theory}, 
Phys. Lett. {\bf B 244}, 20 (1990).

\bibitem{ToppanZhang} F. Toppan, Y.-Z. Zhang, 
\textit{Superconformal affine Liouville theory},
Phys. Lett. B \textbf{292}, 67 (1992).

\bibitem{ChaiKu} M. Chaichian and P. P. Kulish, 
\textit{On the method of inverse scattering problem and B\"acklund transformations for supersymmetric equations}, 
Phys. Lett. B \textbf{78}, 413 (1978).

\bibitem{LeZSavLei} A. N. Leznov, M. V. Saveliev and D. A. Leites, 
\textit{Superalgebra $b(0,1)$ and explicit integration of the supersymmetric Liouville equation}, 
Phys. Lett. B \textbf{96}, 97 (1980).

\bibitem{Arvis} J. F. Arvis, 
	\textit{Classical dynamics of the supersymmetric Liouville theory}, 
	Nucl. Phys. B \textbf{212}, 151 (1983).


\bibitem{ivkr} E. A. Ivanov and S. O. Krivonos, \textit{$U(1)$-supersymmetric extension of the Liouville equation},
Lett. Math. Phys. {\bf 7}, 523 (1983).

\bibitem{ToppanSL} F. Toppan, 
\textit{On the superLiouville theory}, 
Phys. Lett. B \textbf{260}, 346 (1991).

\bibitem{yaji} W. M. Yang and S. C. Jing, {\it A new kind of graded Lie algebra and parastatistical supersymmetry}, Sci. in China (Series A) {\bf 44}, 9 (2001).

\bibitem{tol1} V. N. Tolstoy, {\it Once more on parastatistics}, {Phys. Part. Nucl. Lett.} {\bf{11}},  933 (2014).

\bibitem{StoVDJ} N. I. Stoilova, J. Van der Jeugt, 
\textit{The $\mathbb{Z}_2 \times \mathbb{Z}_2$-graded Lie superalgebra $ \mathfrak{pso}(2m+1|2n) $ and new parastatistics representations}, 
J. Phys. A: Math. Theor. \textbf{51}, 135201 (2018).

\bibitem{StoVDJ3} N. I. Stoilova, J. Van der Jeugt, 
\textit{The $\mathbb{Z}_2 \times \mathbb{Z}_2$-graded Lie superalgebras $\mathfrak{pso}(2n+1|2n)$ and $ \mathfrak{pso}(\infty|\infty)$, and parastatistics Fock spaces}, 
J. Phys. A: Math. Theor. \textbf{55},  045201 (2022).

\bibitem{StoVDJosp} N. I. Stoilova, J. Van der Jeugt, 
\textit{Orthosymplectic $ \mathbb{Z}_2 \times \mathbb{Z}_2$-graded Lie superalgebras and parastatistics}, 
J. Phys. A: Math. Theor. \textbf{57}, 095202 (2024).


\bibitem{top2} F. Toppan, {\it Inequivalent quantizations from gradings and $\mathbb{Z}_{2}\times \mathbb{Z}_{2}$-graded parabosons}, J. Phys. A: Math. Theor.  {\bf 54}, 355202 (2021).

\bibitem{top1} F. Toppan, {\it $\mathbb{Z}_{2}\times \mathbb{Z}_{2}$-graded parastatistics in multiparticle quantum Hamiltonians}, J. Phys. A: Math. Theor. {\bf 54}, 115203 (2021).

\bibitem{bfrt} M. M. Balbino, I. P. de Freitas, R. G. Rana and F. Toppan, \textit{Inequivalent $Z_2^n$-graded brackets, $n$-bit parastatistics and statistical transmutations of supersymmetric quantum mechanics}, Nucl. Phys. \textbf{B 1009}, 116729 (2024).

\bibitem{top3} F. Toppan, {\textit{Transmuted spectrum-generating algebras and detectable parastatistics of the superconformal quantum mechanics}}, J. Phys.: Conf. Ser. {\bf 2667}, 012014 (2023).

\bibitem{WaHa} Z. Wang and K. R. A. Hazzard, 
\textit{Particle exchange statistics beyond fermions and bosons}, 
Nature \textbf{637}, 314 (2025). 


\bibitem{parasim} C. Huerta Alderete and B. M. Rodr\'\i guez-Lara, {\it Quantum simulation of driven para-Bose oscillators}, Phys. Rev. {\bf A 95}, 013820 (2017). 

\bibitem{paraexp} C. Huerta Alderete, A. M. Greene, N. H. Nguyen, Y. Zhu, B. M. Rodr\'\i guez-Lara and N. M. Linke,
{\it Experimental realization of para-particle oscillators}, Scientific Reports, \textbf{15}, 41498 (2025).

\bibitem{polyak} A. M. Polyakov, 
\textit{Gauge transformations and diffeomorphisms}, Int. J. Mod. Phys. A {\bf  5},
833 (1990).

\bibitem{LezSav1} A. N. Leznov and M. V. Savel'ev, 
\textit{Representation of zero curvature for the system of non-linear partial differential equations $ x_{\alpha,z\bar{z}} = \exp(kx)_{\alpha}$ and its integrability}, 
Lett. Math. Phys. \textbf{3}, 489 (1979).

\bibitem{LezSav2} A. N. Leznov and M. V. Savel'ev, 
\textit{Two-dimensional exactly and completely integrable dynamical systems: Monopoles, instantons, dual models, relativistic strings, lund-regge model, generalized Toda lattice, etc.}, 
Comm. Math. Phys. \textbf{89}, 59 (1983).

\bibitem{LezSav3} A. N. Leznov and M. V. Savel'ev, 
\textit{Two-dimensional supersymmetric nonlinear equations associated with embeddings of the superalgebra $osp(1,2)$ in Lie superalgebras}, 
Theor. Math. Phys. \textbf{61}, 1056 (1984).


\bibitem{zhe} A. A. Zheltukhin, 
\textit{Para-Grassmann extension of the Neveu-Schwartz-Ramond algebra}, Theor. Math. Phys. {\bf 71},  491  (1987) (Teor. Mat. Fiz. {\bf 71},  218 (1987)).

\bibitem{fss} L. Frappat, A. Sciarrino and P. Sorba,  \textit{Structure of Basic Lie Superalgebras and of Their Affine Extensions}, Comm. Math. Phys. \textbf{121}, 457 (1989).

\bibitem{vritt} V. Rittenberg, ``{A Guide to Lie Superalgebras}" in \textit{Lecture Notes in Physics} \textbf{79} p. 3; editors P. Kramer and A. Rickers, Springer-Verlag, Berlin (1978).

\bibitem{ivto} E. Ivanov and F. Toppan, \textit{N=2 Superconformal affine Liouville theory},  Phys. Lett. {\bf B 309}, 289 (1993).

\bibitem{brusigma} A. J. Bruce, \textit{${\mathbb Z}_2\times{\mathbb Z}_2$-graded supersymmetry: 2-d sigma models},  J. Phys. A: Math. Theor. {\bf 53}, 455201 (2020).

\bibitem{bruSG} A. J. Bruce, \textit{Is the $\mathbb{Z}_2 \times \mathbb{Z}_2$-graded sine-Gordon equation integrable?}, Nucl. Phys. B {\bf 971}, 115514 (2021).

\bibitem{AiFujiIto} N. Aizawa, I. Fujii and R. Ito, 
\textit{An integrable hierarchy associated with loop extension of $\mathbb{Z}_2^2 $-graded $\mathfrak{osp}(1|2)$},
J. Phys. A:Math. Theor. \textbf{59}, 215206 (2026).

\bibitem{AiFujiItoVas} N. Aizawa, I. Fujii, R. Ito, L. Vasconcellos, 
\textit{An integrable $\mathbb{Z}_2^2$-graded extension of Camassa-Holm equation and its bi-Hamiltonian structure}, 
arXiv:2609.19543 [nlin.SI]. 


\bibitem{yuan2026super} B. Yuan, Y. Bi, Y. Ruan, T. Zhang, 
\textit{Super integrable hierarchies associated with color Lie algebra}, 
arXiv:2605.27897 [nlin.SI]. 


\bibitem{stvdjclass} N. I. Stoilova and J. Van der Jeugt, 
{\textit{On classical $\mathbb{Z}_2\times \mathbb{Z}_2$-graded Lie algebras}}, 
J. Math. Phys. {\bf 64}, 061702 (2023).

\bibitem{ItoNagoTani} R. Ito, A. Nago and S. Tanigawa, 
\textit{On generalized statistics and stability in $\mathbb{Z}_2^2$-graded supersymmetric Yang-Mills theory}, 
Nucl. Phys. B \textbf{1031}, 117654 (2026). 


\bibitem{ItoNago} R. Ito and A. Nago, 
\textit{Novel possible symmetries of S-matrix generated by $\mathbb{Z}_2^n$-graded Lie superalgebras}, 
Nucl. Phys. B \textbf{1014}, 116877 (2025).



\bibitem{sab1} W. A. Sabra, \textit{Hidden Kac-Moody symmetry and 2D Quantum Supergravity},  Nucl. Phys. {\bf B 375}, 82 (1992).

\bibitem{sab2} W. A. Sabra, \textit{Classical Hamiltonian Reduction and Superconformal Algebras}, Phys. Lett. {\bf B 313}, 68 (1993).


\bibitem{Ree} R. Ree, 
\textit{Generalized Lie elements}, Canad. J. Math. \textbf{12}, 493 (1960).

\bibitem{sch} M. Scheunert, 
\textit{Generalized Lie algebras}, 
J. Math. Phys. {\bf 20}, 712 (1979).

\bibitem{GrJa} H. S. Green and P. D. Jarvis,  1983 \textit{Casimir invariants, characteristic identities, and Young diagrams for color algebras and superalgebras},  J. Math. Phys. \textbf{24},  1681 (1983). 


\bibitem{AAD} N. Aizawa, K. Amakawa, S. Doi, 
\textit{$\mathcal{N}$-Extension of double-graded supersymmetric and superconformal quantum mechanics}, 
J. Phys. A: Math. Theor. \textbf{53}, 065205 (2020).

\bibitem{RY} R. Liu and Y. Tan, 
\textit{Construction of color Lie algebras from homomorphism of modules of Lie algebras}, 
J. Algebra \textbf{620} (2023), 1.

\bibitem{aizt} N. Aizawa, R. Ito, Z. Kuznetsova and F. Toppan, \textit{New aspects of the $\Z2$-graded $1D$ superspace: induced strings and $2D$ relativistic models}, 
Nucl. Phys. {\bf B 991}, 116202 (2023). 

\bibitem{IvaKri} E. A. Ivanov and S. O. Krivonos, 
\textit{$U(1)$-supersymmetric extension of the Liouville quations}, 
Lett. Math. Phys. \textbf{7}, 523 (1983).

\bibitem{Anomaly} F. Toppan, \textit{On anomalies in classical dynamical systems}, J. Nonlinear Math. Phys. \textbf{8}, 518 (2001).

\bibitem{Ramond} P. Ramond, \textit{Dual Theory for Free Fermions}, Phys. Rev. {\bf D 3}, 2415 (1971).

\bibitem{NS} A. Neveu and J. H. Schwarz, \textit{Factorizable dual model of pions},  Nucl.Phys. {\bf B 31},  86 (1971).

\bibitem{NAJS} N. Aizawa and J. Segar, \textit{Affine extensions of $\Z2$-graded $osp(1|2)$ and Virasoro algebra}, Int. J. Geom. Methods Mod. Phys. \textbf{23}, 2540052 (2026). 

\bibitem{MRY} R. V. Moody, S. E. Rao and T. Yokonuma, 
\textit{Toroidal Lie algebras and vertex representations}, 
Geom. Dedicata \textbf{35}, 283 (1990).

\bibitem{afgz} H. Aratyn, L. A. Ferreira, J. F. Gomes and  A. H. Zimerman, \textit{Kac-Moody construction of Toda type field theories}, Phys. Lett. {\bf B 254}, 372 (1991).

\bibitem{NARITT2D} N. Aizawa, Ren Ito and T. Tanaka, 
\textit{$\mathbb{Z}_2^2$-graded supersymmetry via superfield on minimal $\mathbb{Z}_2^2$-superspace}, 
J. Phys. A: Math. Theor. \textbf{57}, 435201 (2024).

\end{thebibliography}
\end{document}